\documentclass[structabstract]{aa}  
\usepackage{graphicx}
\usepackage{txfonts}
\usepackage{natbib} 
\bibpunct{(}{)}{;}{a}{}{,} 
\newcommand{\micron}{\mu{\rm m}}
\usepackage{color}
\begin{document}
   \title{AKARI/IRC $18~\micron$ Survey of Warm Debris Disks}

   \authorrunning{Fujiwara et al.}


   \author{Hideaki~Fujiwara \inst{1},
Daisuke~Ishihara \inst{2},
Takashi~Onaka\inst{3}, 
Satoshi~Takita\inst{4}, 
Hirokazu~Kataza\inst{4}, 
Takuya~Yamashita\inst{5}, 
Misato~Fukagawa\inst{6}, 
Takafumi~Ootsubo\inst{7}, 
Takanori~Hirao\inst{8}, 
Keigo~Enya\inst{4},
Jonathan~P.~Marshall\inst{9}, 
Glenn~J.~White\inst{10,11},
Takao~Nakagawa\inst{4}, 
\and 
Hiroshi~Murakami\inst{4} 
          }

   \institute{Subaru Telescope, National Astronomical Observatory of Japan, 
650 North A'ohoku Place, Hilo, HI 96720, USA\\
              \email{hideaki@naoj.org}
         \and
             Graduate School of Science, Nagoya University, Furo-cho, Chikusa-ku, Nagoya, Aichi 464-8602, Japan
         \and
Department of Astronomy, School of Science, University of Tokyo, Bunkyo-ku, Tokyo 113-0033, Japan
         \and
             Institute of Space and Astronautical Science, Japan Aerospace Exploration Agency, 
3-1-1 Yoshinodai, Chuo-ku, Sagamihara, Kanagawa 252-5210, Japan
         \and
National Astronomical Observatory of Japan, 2-21-1 Osawa, Mitaka, Tokyo 181-0015, Japan
         \and
Graduate School of Science, Osaka University, 1-1 Machikaneyama, Toyonaka, Osaka 560-0043, Japan
         \and
Astronomical Institute, Tohoku University, 6-3 Aramaki, Aoba-ku, Sendai, Miyagi 980-8578, Japan
         \and
Research Institute of Science and Technology for Society, Japan Science and Technology Agency, K's Gobancho Bldg, 7, Gobancho, Chiyoda-ku, Tokyo 102-0076, JAPAN
         \and
Departmento F\'{i}sica Te\'{o}rica, Facultad de Ciencias, Universidad Aut\'{o}noma de Madrid, Cantoblanco, 28049 Madrid, Spain 
         \and
Department of Physics and Astronomy, The Open University, Walton Hall, Milton Keynes, MK7 6AA, UK
         \and
Space Science \& Technology Department, The Rutherford Appleton Laboratory, Chilton, Didcot, Oxfordshire OX11 0QX, UK
             }

   \date{Received 19 June 2012 / Accepted 21 November 2012}

 
  \abstract
{Little is known about the properties of the warm ($T_{\rm dust} \gtrsim 150$~K) debris disk material 
located close to the central star, which has a more direct link to the formation of terrestrial planets 
than the low temperature debris dust that has been detected to date.}
{To discover new warm debris disk candidates that show large $18~\micron$ excess and estimate the fraction of stars with excess 
based on the {\it AKARI}/IRC Mid-Infrared All-Sky Survey data.}
{We have searched for point sources detected in the {\it AKARI}/IRC All-Sky Survey, 
which show a positional match with A-M dwarf stars in the Tycho-2 Spectral Type Catalogue
and exhibit excess emission at $18~\micron$ compared to that expected from the $K_{\rm S}$ magnitude in the 2MASS catalogue. }
{We find 24 warm debris candidates including 8 new candidates among A-K stars. 
The apparent debris disk frequency is estimated to be $2.8 \pm 0.6$\%. 
We also find that A stars and solar-type FGK stars have
different characteristics of the inner component of the identified
debris disk candidates --- while debris disks around A stars are cooler
and consistent with steady-state evolutionary model of debris disks,
those around FGK stars tend to be warmer and cannot be explained by
the steady-state model.}
   {}


   \maketitle
%

\section{Introduction}

Some main-sequence stars are known to have dust disks around them. 
Primordial dust grains that originally exist in protoplanetary disks 
around T Tauri stars or Herbig Ae/Be stars are dissipated 
before their central stars reach the main-sequence phase. 
Therefore the circumstellar dust grains around main-sequence stars 
should be composed of second generation dust grains replenished during the main-sequence phase, 
rather than primordial dust from protoplanetary disks.  
These second generation dust grains are thought to have originated from collisions of planetesimals, 
or during the destruction of cometary objects \citep[e.g.][]{backman93,lecavelier96}, 
giving a reason why circumstellar dust disks around main-sequence stars 
are named ``debris disks.''
Debris disks are expected to be related to the
stability of minor bodies, and potentially the presence of planets around
stars \citep[e.g.][]{wyatt08}. 

Debris disks are identified from the spectral energy distributions (SEDs) of
stars that show an excess over
their expected photospheric emission at infrared (IR) wavelengths, 
since circumstellar dust grains absorb the stellar light and re-emit mainly in the IR region. 
The first example of the debris disk, around Vega, 
was discovered through its excess emission at wavelengths of $\gtrsim 60~\micron$  
by {\it Infrared Astronomical Satellite} ({\it IRAS}) in 1980s \citep{aumann84}. 
After the discovery of the Vega debris disk, more than 100 other debris disks were 
identified from the {\it IRAS} catalogue. 
A number of new debris disk candidates have been discovered recently, 
even though the {\it IRAS} mission flew more than 20 years ago \citep{rhee07}. 
The power of a next generation all-sky survey satellite mission like {\it AKARI} \citep{murakami07} to take
forward the debris disk search is evident.

Most of the known debris disks only show excess far-infrared (FIR) emission at wavelengths longer than $25~\micron$. 
This excess comes from the thermal emission of dust grains with low temperatures 
($T_{\rm dust} \sim 100$~K), and is an analogue of Kuiper belt objects in the Solar System. 
Recently {\it Herschel} \citep{pilbratt10} has provided FIR data for debris disks with high sensitivity and high spatial resolution 
and is opening a new horizon of research on cold debris disks \cite[e.g.][]{thompson10}.
For example, \cite{eiroa11} discovered cold ($T_{\rm dust} \lesssim 20$~K), faint debris disks towards three GK stars, 
which might be in a new class of coldest and faintest disks discovered so far around mature stars 
and which cannot easily be explained by invoking classical debris disk models. 
In addition, some cold debris disks have been spatially resolved by {\it Herschel} \citep[e.g.][]{loehne12,liseau10}. 
On the other hand, 
little is known to date about the warm ($T_{\rm dust} \gtrsim 150$~K) debris disk material located close to the star, 
which should be an analogue of the asteroid belt in the Solar System. 
Warm dust grains in the inner region of debris disks 
should have a more direct link to the formation of terrestrial planets 
than the low temperature dust that has been previously studied \citep{meyer08}.
Recent high-sensitivity surveys at $10-20~\micron$ allow us to investigate  
the properties of this inner debris disk material. 

{\it AKARI} is a Japanese IR satellite dedicated primarily 
to an IR all-sky survey \citep{murakami07}. It was launched in February 2006.
The Mid-Infrared (MIR) All-Sky Survey was performed using 9 and $18~\micron$ broad band filters 
with the InfraRed Camera \citep[IRC;][]{onaka07} onboard {\it AKARI} until August 2008 \citep{ishihara10}, 
and a complementary FIR survey at 65--160$~\micron$ was performed 
with the Far-Infrared Surveyor \citep{yamamura10}. 
We show that the data from the {\it AKARI}/IRC All-Sky Survey 
is very powerful to identify warm debris disks and to contribute to 
increase the number of sample with its higher sensitivity and spatial resolution 
than those of {\it IRAS}.

Here we report initial results of survey of warm debris disks around main-sequence stars 
based on the {\it AKARI}/IRC All-Sky Survey.


\section{$18~\micron$ Excess Search with {\it AKARI}/IRC All-Sky Survey}

\subsection{Input Catalogue: Tycho-2 Spectral Type Catalogue}

In this work, we use the Tycho-2 Spectral Type Catalogue \citep{wright03} as an input catalogue. 
The Tycho-2 Spectral Type Catalogue lists spectral types for 351864 of the Tycho-2 stars 
by cross-referencing Tycho-2 to several catalogues that have spectral types (the Michigan Catalogue for the HD Stars, 
Volumes 1-5 \citep{houk75,houk78,houk82,houk88,houk99}; 
the Catalogue of Stellar Spectra Classified in the Morgan-Keenan System \citep{jaschek64}; 
the MK Classification Extension \citep{kennedy96}; the Fifth Fundamental Catalogue (FK5) Part I \citep{fricke88} 
and Part II \citep{fricke91}; and the Catalogue of Positions and Proper Motions (PPM), 
North \citep{roeser88} and South \citep{bastian93}), using the VizieR astronomical database\footnote{URL: http://vizier.u-strasbg.fr/viz-bin/VizieR/.}.

We selected stars whose luminosity classes were labeled as ``V'' (dwarf) from the catalogue to remove giant stars from the sample, 
since we focus our search for debris disks around main-sequence stars. 
Since young and massive main-sequence stars, which are commonly known as classical Be stars, 
sometimes exhibit significant IR excess resulting from free-free emission from circumstellar gas disks,  
stars whose spectral types were labelled as ``O'' or ``B'' in the Tycho-2 Spectral  Type Catalogue were excluded from our sample. 
The selection of A-M dwarfs from the Tycho-2 Spectral Type Catalogue left 
64209 stars as an input catalogue for further investigation. 
Since the $V$-band limiting magnitude in the Tycho-2 Catalogue is about $V = 11.5$ (Hog et al. 2000), 
the distance limit for A0, F0, G0, K0, and M0 dwarfs probed by the Tycho-2 Catalogue 
is estimated as 1500, 580, 260, 130, and 35~pc, respectively, by assuming no extinction at the $V$-band.

\begin{figure}
\centering
\includegraphics[height=4.7cm, width=7.5cm]{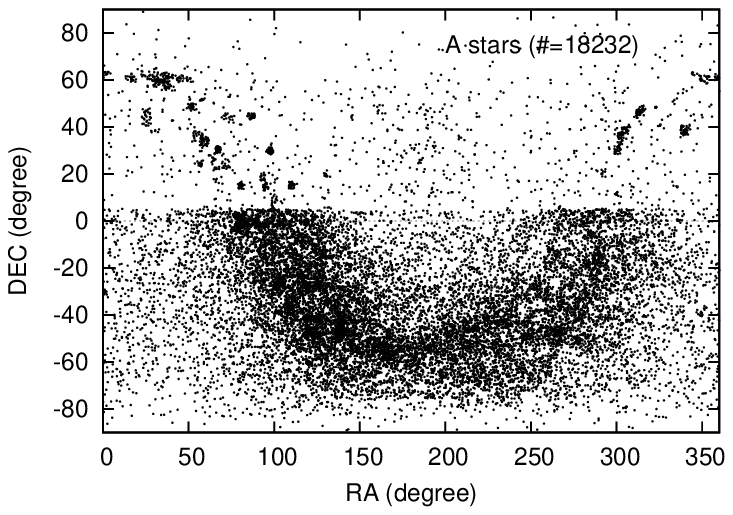}
\includegraphics[height=4.7cm, width=7.5cm]{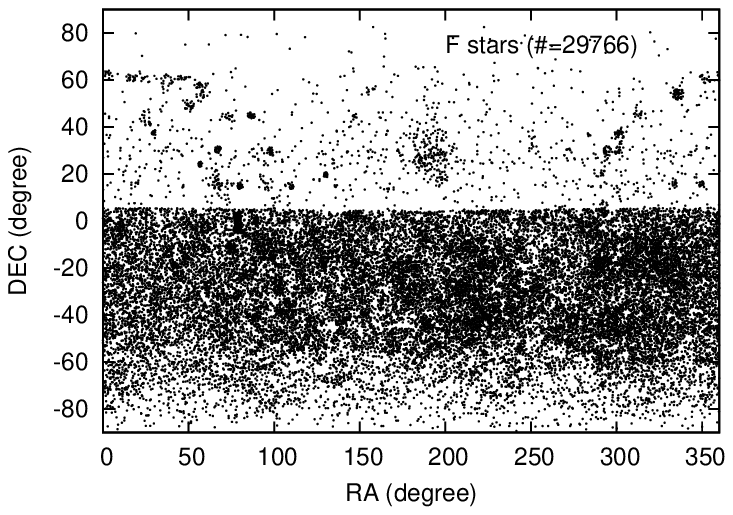}
\includegraphics[height=4.7cm, width=7.5cm]{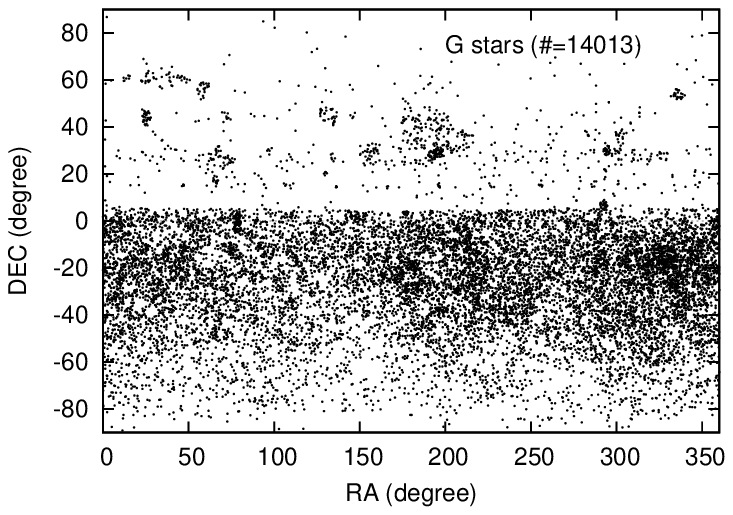}
\includegraphics[height=4.7cm, width=7.5cm]{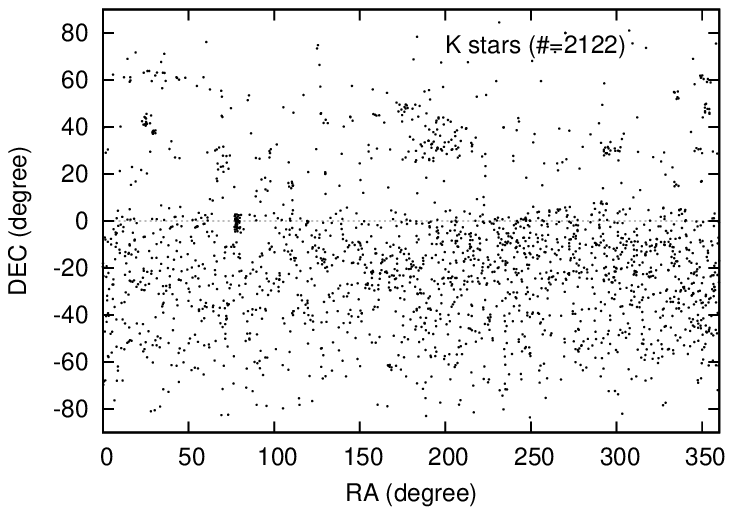}
\includegraphics[height=4.7cm, width=7.5cm]{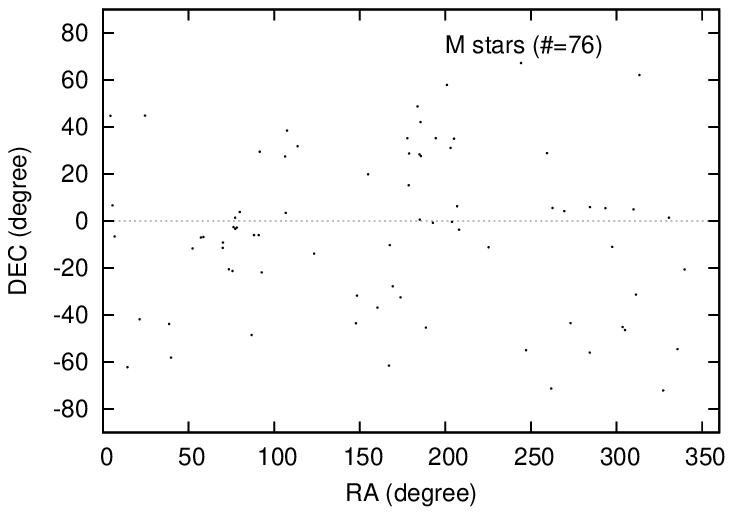}
\caption{Spatial distributions of A-M dwarf stars from the Tycho-2 Spectral Type Catalogue \citep{wright03}.
\label{T2Sdist}}
\end{figure}

The spatial distributions of the A-M dwarf stars from the Tycho-2 Spectral Type Catalogue 
in equatorial coordinates are shown in Figure~\ref{T2Sdist}. 
Since the information on luminosity classes in the Tycho-2 Spectral Type Catalogue has been adopted mainly from the Michigan Catalogue, 
which was based on the observations taken at Cerro Tololo Inter-American Observatory with the Michigan Curtis Schmidt telescope, 
the spatial distribution of the A-M dwarf stars used in this study is unevenly distributed in the southern hemisphere.

\subsection{AKARI/IRC Point Source Catalogue}

In the {\it AKARI} All-Sky Survey, the IRC was operated in the scan mode with a scan speed of $215\arcsec$ s$^{-1}$ 
and the data sampling time of 0.044 s, which provided a spatial sampling of $\sim 9\farcs4$ along the scan direction. 
The pixel scale along the cross-scan direction was $\sim 9\farcs4$
by binning 4 pixels in a row to reduce the data downlink rate \citep{ishihara10}. 
The positional accuracy is improved to $\sim 5\arcsec$ in both bands by combining the dithered data 
from multiple observations using the standard {\it AKARI}/IRC All-Sky Survey pipeline software \citep{ishihara10}. 
The $5~\sigma$ sensitivity for a point source per scan is estimated to be 50 mJy in the S9W band and 90 mJy in the L18W band. 
The distance limit at which the stellar photosphere can be detected 
in the L18W band of the {\it AKARI}/IRC Point Source Catalogue (PSC) is estimated as 74, 40, 26, 17, and 10~pc for A0, F0, G0, K0, and M0 dwarfs, respectively.

The first version of the {\it AKARI}/IRC PSC contains 
more than 850000 sources detected at $9~\micron$ and 195000 sources at $18~\micron$. 
The quoted flux density and its uncertainty for a single source were calculated 
as average and deviation (root-mean-square) of photometry of multiple detections. 
Details of the {\it AKARI}/IRC All-Sky Survey and its data reduction process are described 
in \citet{ishihara10}.

\subsection{Cross-correlation}

The selected dwarfs from the Tycho-2 Spectral Type Catalogue were cross-correlated against the {\it AKARI}/IRC PSC positions. 
A search radius of $5\arcsec$ was adopted to take account of the positional uncertainty in the {\it AKARI}/IRC PSC. 
After the results of the cross-correlation, 873 stars out of the input 64209 stars were detected at $18~\micron$. 
For the detected stars, we searched for the nearest near-infrared (NIR) counterparts within a search radius of $5\arcsec$ in 
the 2MASS All-Sky Catalogue of Point Sources \citep{cutri03} 
and found 856 stars with 2MASS counterparts. 
The numbers of stars that were detected with both of {\it AKARI}/IRC $18~\micron$ and 2MASS are summarised in Table~\ref{frequency}.


\begin{table*}
\caption{Summary of debris disk survey at $18~\micron$. \label{frequency}}
\centering                          
\begin{tabular}{cccccccccc}
\hline\hline                 
Spectral &  Input   & $18~\micron$-Detected   &  Excess   &  Debris           &  Apparent  & $18~\micron$ Sources & Debris Candidates   & Debris Disk \\
Type     &  Sources &  Sources                           &  Sources  &  Candidates   &  Frequency & w/ Detectable              & w/ Detectable          & Frequency   \\    
            &                &                                      &                &                      &  (\%)           & Photoshpere                &  Photoshpere           & (\%)             \\    
\hline                        
A   &  18232   & 196         &  21      &  11      &  $5.6 \pm 1.6$     & 178 & 4 & $2.1 \pm 1.1$ \\
F   &  29766   & 324         &  12      &  10      &  $3.1 \pm 1.0$     & 311 & 3 & $1.0 \pm 0.6$ \\
G   &  14013   & 173         &   3      &   2      &  $1.2 \pm 0.8$     & 169 & 2 & $1.2 \pm 0.8$ \\
K   &  2122    & 144         &   2      &   1      &  $0.7 \pm 0.7$     & 144 & 0 & 0.0 \\
M   &  76      &  19         &   4      &   0      &  0.0     & 19 & 0 & 0.0 \\
\hline                                   
Total& 64209   & 856         &  42      &  24      &  $2.8 \pm 0.6$     & 830 & 9 & $1.1 \pm 0.4$ \\
\hline                                   
\end{tabular}
\end{table*}

\subsection{$18~\micron$ Excess Candidate Selection}

\begin{figure*}
\centering
\includegraphics[height=6cm, width=4.5cm, angle=-90]{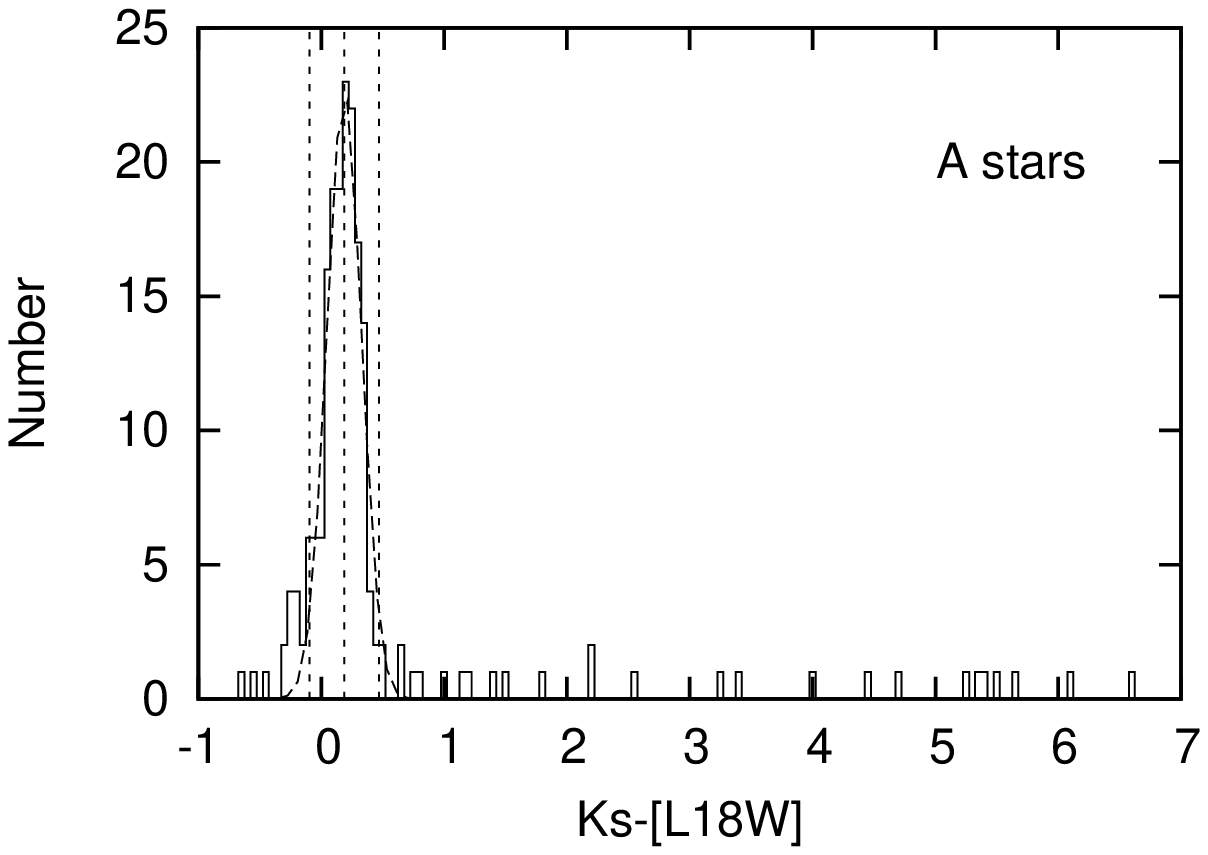}
\includegraphics[height=6cm, width=4.5cm, angle=-90]{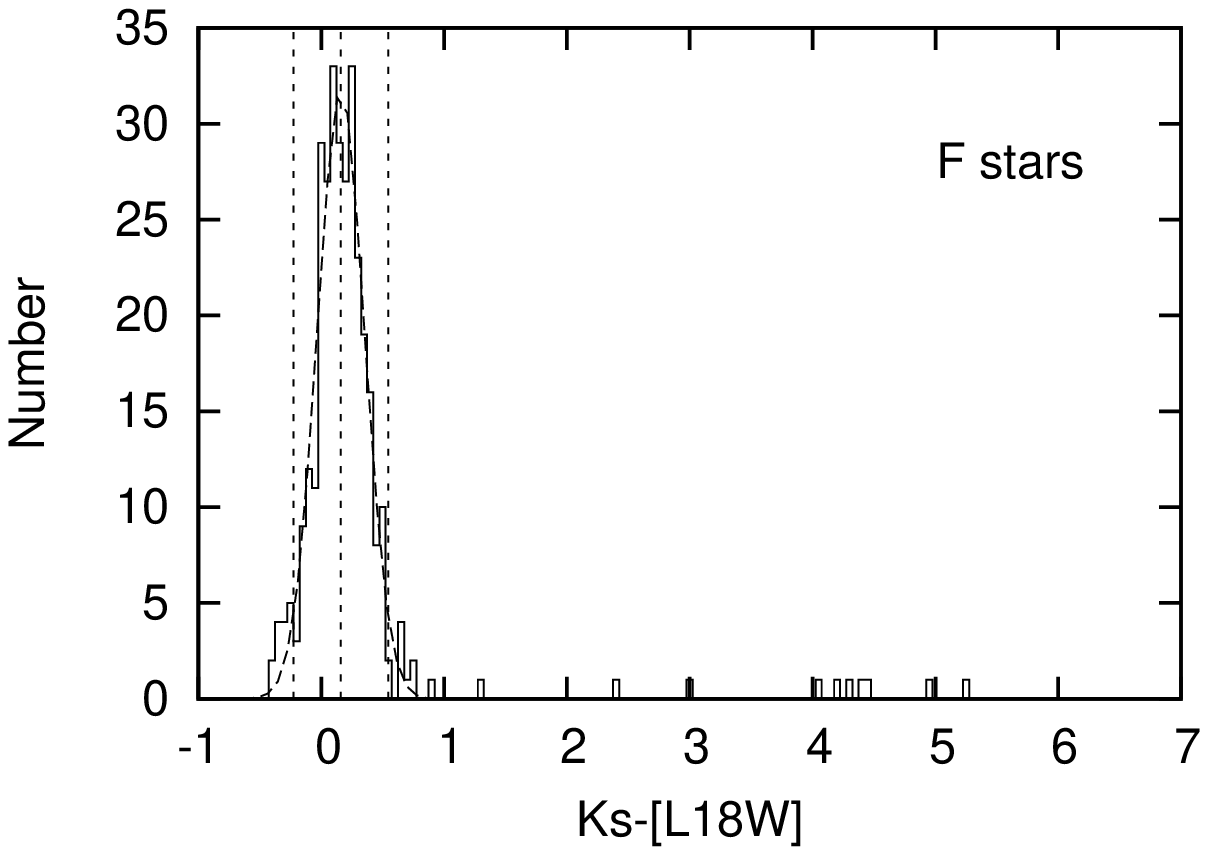}
\includegraphics[height=6cm, width=4.5cm, angle=-90]{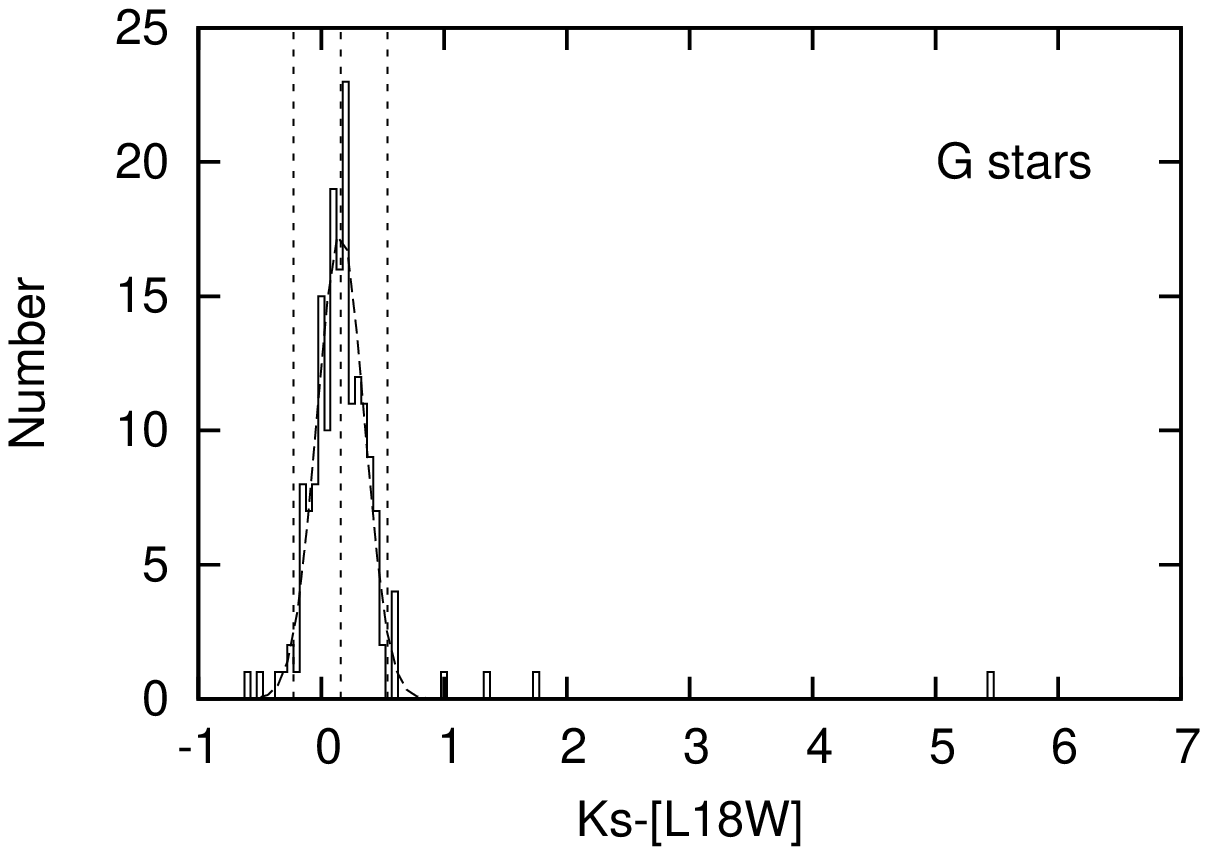}
\includegraphics[height=6cm, width=4.5cm, angle=-90]{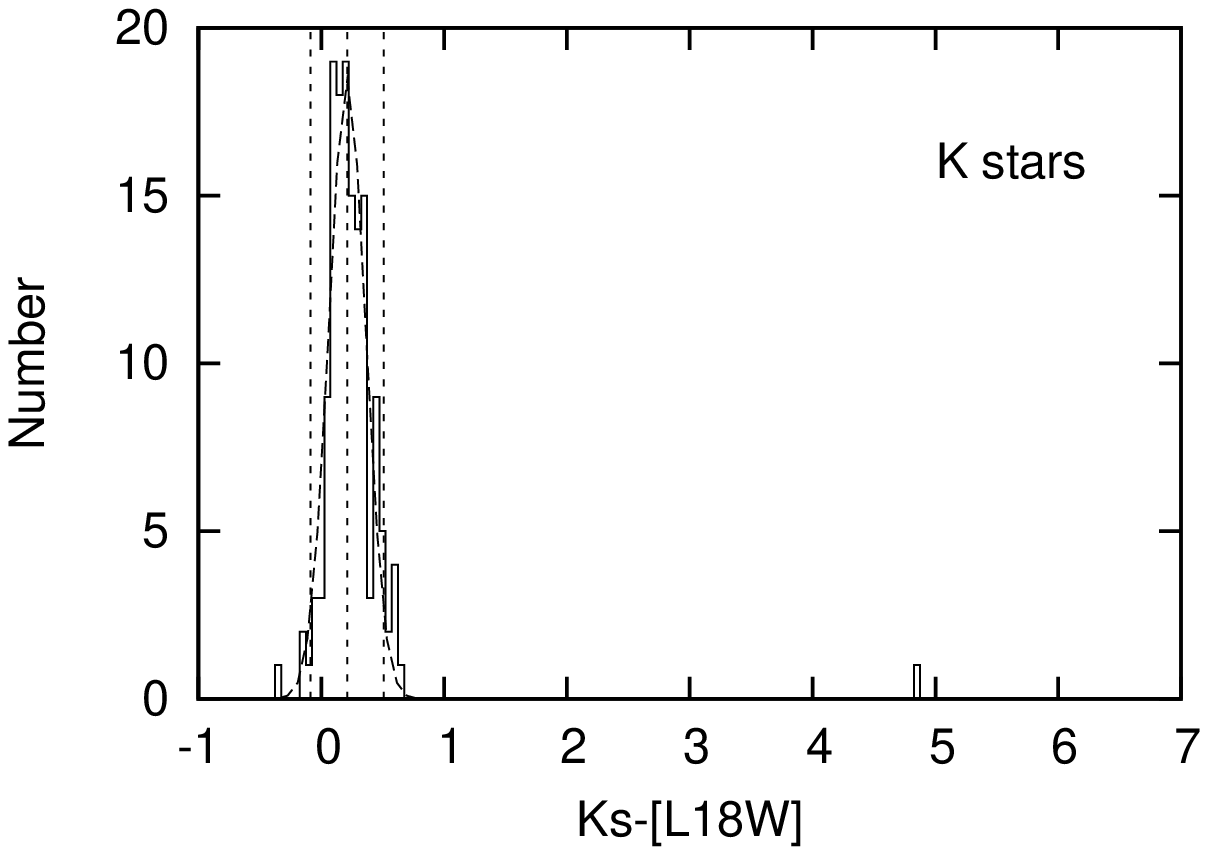}
\includegraphics[height=6cm, width=4.5cm, angle=-90]{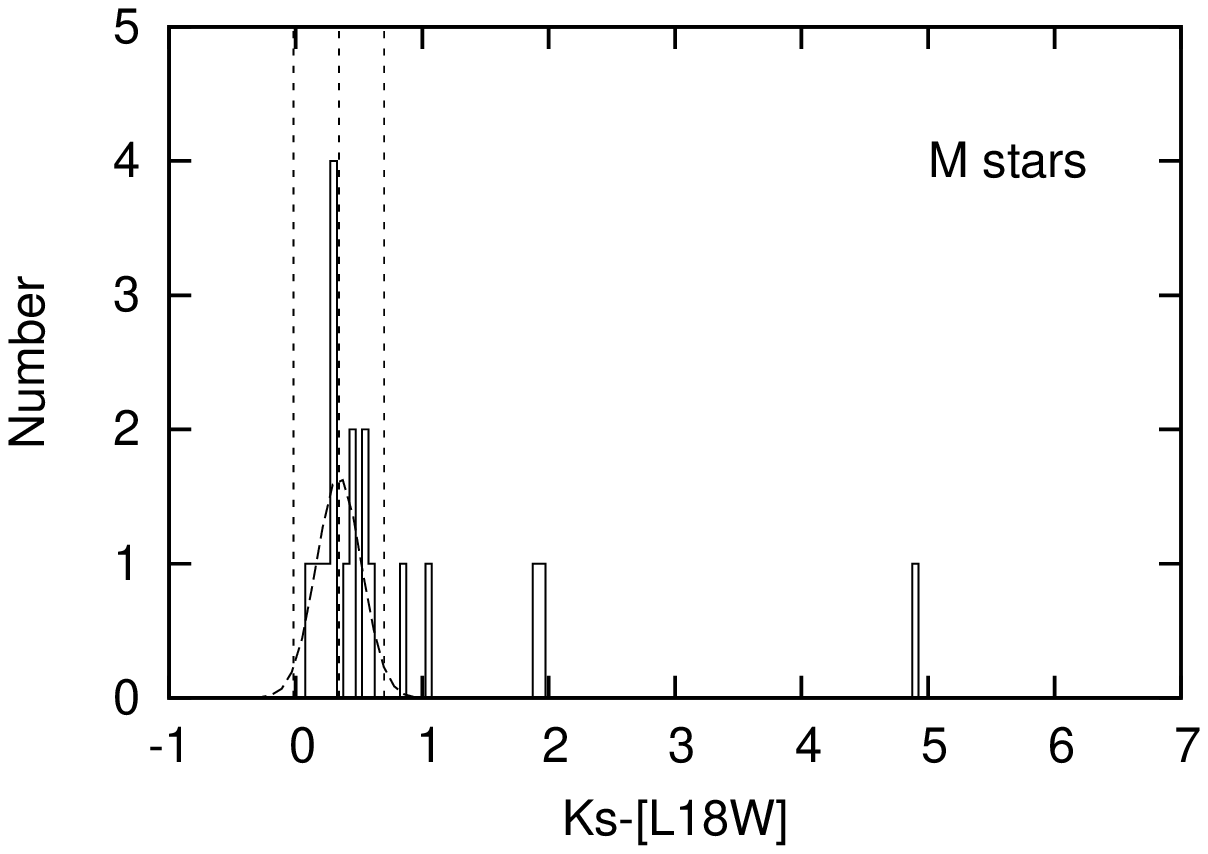}
\caption{Histogram of the $K_{\rm S}-$[18] colour of the stars detected at $18~\micron$ for each spectral type star. 
The results of Gaussian fitting of the histograms are indicated by the dashed lines. 
The three vertical dotted lines indicate the peak position $\mu_{K_{\rm S}-{\rm [L18W]}}$ (middle), 
and $2\sigma$ positions (left and right) of the distribution of the $K_{\rm S}-$[18] colour. 
\label{hist}}
\end{figure*}

We made a histogram of the $K_{\rm S}-$[18] colours of the stars detected at $18~\micron$ for each spectral type 
and set the $K_{\rm S}-$[18] colours as an indicator of $18~\micron$ excess. 
The histograms of the $K_{\rm S}-$[18] colour are shown in Figure~\ref{hist} 
for each spectral type. 
Gaussian fitting was performed for the distribution of the $K_{\rm S}-$[18] colour 
to determine the thresholds for $18~\micron$ excess detection. 
The peak position $\mu_{K_{\rm S}-{\rm [L18W]}}$ and profile width $\sigma_{K_{\rm S}-{\rm [L18W]}}$ 
of the $K_{\rm S}-$[18] colour in each of the spectral types derived by Gaussian fitting are listed in Table~\ref{fitparam}. 
The peak position ($\mu_{K_{\rm S}-{\rm [L18W]}}$) corresponds to the empirical $K_{\rm S}-$[18] colours of normal stars. 
The scatter in the $K_{\rm S}-$[18] colours ($\sigma_{K_{\rm S}-{\rm [L18W]}}$) are between 0.14 and 0.19 mag. 
This scatter can be accounted for by the photometric uncertainties in {\it AKARI}/IRC and 2MASS. 
For faint stars with $K_{\rm S} \gtrsim 4$ mag, the scatter is dominated 
by the uncertainty in the {\it AKARI}/IRC L18W band photometry, 
which is mainly by the background noise ($< 0.2$ mag), 
while the uncertainty in the 2MASS $K_{\rm S}$-band photometry is smaller than 0.02 mag and does not contribute significantly.
For bright stars with $K_{\rm S} \lesssim 4$ mag, the uncertainty in the {\it AKARI}/IRC L18W band photometry is small ($\lesssim 0.05$ mag) 
and the scatter is dominated by the uncertainty in the 2MASS $K_{\rm S}$-band photometry, which is $\sim 0.1-0.2$ mag, 
since those stars are in the range of partial saturation in the 2MASS observations \citep{cutri03}. 

We set a $2\sigma$-threshold for a positive $18~\micron$ excess detection for each spectral type 
as defined in Table~\ref{fitparam}. 
Sources whose $K_{\rm S}-$[18] colour is larger than the $2\sigma$-thresholds are selected 
as $18~\micron$ excess candidates. 
In the selection of the $18~\micron$ excess candidates, 
the photometric uncertainties of the individual stars are also taken into account. 
A visual inspection of the excess candidates was 
conducted by comparing {\it AKARI}/IRC images with 2MASS and DSS images
to confirm that there were no nearby IR sources. 
During visual inspection, HD~68256 and HD~68257, whose separation is $6\arcsec$, 
HD~24071 and HD~24072, whose separation is $4\arcsec$, 
HD~94602 and HD~94601, whose separation is $7\arcsec$, 
and HD~72945 and HD~72946, whose separation is $10\arcsec$, 
and HD~147722 and HD~147723, whose separation is $2\arcsec$, 
were found to be unresolved in the {\it AKARI}/IRC All-Sky image and could be contaminated by each other. 
Thus they were excluded from consideration as $18~\micron$ excess candidates. 
BD+213825 is also found to be contaminated by 2MASS J19334644+2130586, whose separation is $7\arcsec$, 
and thus excluded. 
It should be noted that the contaminating star 2MASS J19334644+2130586 has a significant MIR excess, 
which was confirmed by our follow-up observations using Subaru/COMICS and {\it Spitzer}/IRS. 
The rejected sources due to the nearby star contamination are listed in Table~\ref{reject}.
Although HD~89125 and HD~15407 might be contaminated by the nearby star GJ~387 (separation is $7\arcsec$) 
and HIP~11962 (separation is $20\arcsec$), respectively, 
the observed excess flux densities at $18~\micron$ are much larger that the expected flux densities of the contaminating sources. 
We conclude that the $18~\micron$ excesses of HD~89125 and HD~15407 are secure, 
and we select a final catalogue of 42 stars with $18~\micron$ excess in the Tycho-2 Spectral Type Catalogue. 

The $9~\micron$ flux densities of HD~166191, HD~39415, and HD~145263, 
which were selected from the $K_{\rm S}-$[18] colours, are not available in the latest version 
of {\it AKARI}/IRC PSC since they were observed by {\it AKARI} at $9~\micron$ during a South Atlantic Anomaly (SAA) passage, 
and thus rejected from the PSC. 
However the reliable detections are confirmed in the $9~\micron$ images. 
We measured the $9~\micron$ flux densities in the images manually. 
\cite{mouri11} investigated the behavior of signals from the MIR detector onboard {\it AKARI} in the SAA regions. 
They found that ionising particle hits affected the detector, leading to changes in the output offset. 
Therefore the SAA does not significantly affect measurements of point sources in the {\it AKARI}/IRC All-Sky Survey data, 
and therefore the measured flux densities at $9~\micron$ of HD~166191, HD~39415, and HD~145263 are reliable.

\begin{table}
\begin{minipage}[t]{\columnwidth}
\caption{Obtained Parameters from Gaussian Fit.\label{fitparam}}
\centering
\renewcommand{\footnoterule}{}  
\begin{tabular}{ccccc}
\hline \hline
SpT&  $\mu_{K_{\rm S}-{\rm [L18W]}}$   & $\sigma_{K_{\rm S}-{\rm [L18W]}}$ 
& ${\rm Threshold}(2\sigma)$\footnote{${\rm Threshold}(2\sigma) = \mu_{K_{\rm S}-{\rm [L18W]}}+2\sigma_{K_{\rm S}-{\rm [L18W]}}$.} \\
   & (mag) & (mag) & (mag) \\
\hline
A & 0.187 & 0.142 & 0.470 \\
F & 0.160 & 0.174 & 0.545 \\
G & 0.158 & 0.192 & 0.542 \\
K & 0.212 & 0.149 & 0.511 \\
M & 0.341 & 0.179 & 0.700 \\
\hline
\end{tabular}
\end{minipage}
\end{table}



\begin{table}
\begin{minipage}[t]{\columnwidth}
\caption{List of Rejected Sources. \label{reject}}
\centering
\renewcommand{\footnoterule}{}  
\begin{tabular}{ccc}
\hline \hline
Name & IRC ID & Contamination Source \\
\hline
HD~24071 & 0348356-373717 & HD~24072 \\
HD~94602 & 1055371+244457 & HD~94601 \\
HD~68256 & 0812129+173850 & HD~68257 \\
HD~68257 & 0812129+173850 & HD~68256 \\
HD~72945 & 0835509+063714 & HD~72946 \\
HD~147722 & 1624398-294215 & HD~147723 \\
BD+213825 & 1933467+213057 & 2MASS J19334644+2130586\footnote{2MASS J19334644+2130586 shows IR excess.} \\
\hline
\end{tabular}
\end{minipage}
\end{table}

\subsection{Known YSOs and late-type stars}

Since we focus on debris disks in this paper, we need to distinguish debris disks from other types of 
excess emission such as those often seen in young stellar objects (YSOs) or late-type stars.
We searched for the attributes of the 42 stars exhibiting $18~\micron$ excess in the SIMBAD database\footnote{URL: http://simbad.u-strasbg.fr/simbad/.} 
and find that 12 and 6 stars were classified as YSOs and late-type
stars in the literature, respectively. 
The known YSOs and late-type stars in the $18~\micron$ excess sample are listed in Table~\ref{yso}. 
The spectral type and luminosity classes in Table~\ref{yso} were taken from the SIMBAD database, 
which is supposed to contain the latest information. 
For stars whose stellar information was revised after the Tycho-2 Spectral Type Catalogue, 
the spectral type and luminosity class listed in Table~\ref{yso} may differ from those in the Tycho-2 Spectral Type Catalogue. 
Thus stars other than luminosity class V have been included. 
24 main-sequence stars remain as candidates for debris disks after excluding the known YSOs and late-type stars.


\begin{table*}
\begin{minipage}[t]{\columnwidth}
\caption{List of known YSOs and late-type stars detected at $18~\micron$. \label{yso}}
\centering                          
\renewcommand{\footnoterule}{}  
\begin{tabular}{ccccccccc}
\hline \hline
Name & IRC ID & $F_{{\rm obs},9}$ & $F_{{\rm obs},18}$ & $K_{\rm S}-$[9] & $K_{\rm S}-$[18] &
Spectral & Stellar & Ref.\footnote{References. (1) \cite{montesinos09}; (2) \cite{acke09}; (3) \cite{hernandeez05}; (4) \cite{manoj06}; (5) \cite{abia09}; 
(6) \cite{gielen08}; (7) \cite{schutz09}; (8) \cite{winters03}; (9) \cite{sloan98}; (10) \cite{josselin98}.} \\
 & & (Jy) & (Jy) & (mag) & (mag) & Type & Type & \\
\hline
HD~2326   & 0027064-063616 &  99.943 $\pm$ 0.725 & 58.287 $\pm$ 0.641 & 0.86 & 1.95 & M7     & O-rich AGB  & 8 \\
HD~23937  & 0348475-070053 &  37.028 $\pm$ 0.144 & 13.536 $\pm$ 0.217 & 0.47 & 1.05 & M3     & O-rich AGB  & 9 \\
HD~37258  & 0536592-060916 &   1.749 $\pm$ 0.004 &  1.622 $\pm$ 0.046 & 3.92 & 5.53 & A2V    & Herbig Ae  & 2 \\
HD~37357  & 0537471-064230 &   1.513 $\pm$ 0.001 &  1.671 $\pm$ 0.060 & 3.44 & 5.23 & A0e    & Herbig Ae  & 2 \\
HD~38238  & 0544187+000840 &   1.221 $\pm$ 0.004 &  1.690 $\pm$ 0.059 & 2.69 & 4.72 & A7IIIe & Herbig Ae  & 3 \\
HD~68695  & 0811445-440508 &   0.722 $\pm$ 0.010 &  1.275 $\pm$ 0.061 & 3.35 & 5.65 & A0V    & Herbig Ae  & 4 \\
HD~100453 & 1133055-541928 &   5.518 $\pm$ 0.077 & 18.744 $\pm$ 0.248 & 3.08 & 6.08 & A9Ve   & Herbig Ae  & 2  \\
HD~107439 & 1221125-491240 &   4.724 $\pm$ 0.422 &  3.556 $\pm$ 0.132 & 4.07 & 5.44 & G4Vp   & post-AGB  & 6 \\
HD~120806 & 1351516-034033 &  58.500 $\pm$ 0.627 & 29.671 $\pm$ 0.085 & 0.97 & 1.91 & M      & O-rich AGB  & 8  \\
HD~135344 & 1515484-370915 &   1.727 $\pm$ 0.063 &  3.285 $\pm$ 0.114 & 2.06 & 4.44 & A0V    & Herbig Ae  & 2 \\
HD~139614 & 1540463-422953 &   2.386 $\pm$ 0.057 & 10.568 $\pm$ 0.118 & 3.32 & 6.61 & A7Ve   & Herbig Ae  & 1 \\
HD~142666 & 1556400-220140 &   5.152 $\pm$ 0.355 &  6.578 $\pm$ 0.007 & 3.48 & 5.42 & A8Ve   & Herbig Ae  & 2 \\
HD~144432 & 1606579-274310 & --             &  7.326 $\pm$ 0.259 & -- & 5.35 & A9/F0V & Herbig Ae & 2  \\
HD~152404 & 1654448-365318 &   2.032 $\pm$ 0.025 &  2.870 $\pm$ 0.029 & 2.90 & 4.95 & F5Ve   & T Tauri  & 2 \\
HD~155555 & 1717255-665704 &   0.816 $\pm$ 0.025 &  0.282 $\pm$ 0.027 & 0.11 & 0.63 & G7IV   & T Tauri  & 7 \\
HD~157045 & 1727235-711715 &  80.030 $\pm$ 0.106 & 85.198 $\pm$ 0.310 & 3.18 & 4.92 & M3III  & O-rich AGB  & 10 \\
HD~158643 & 1731250-235745 & --             &  9.241 $\pm$ 0.111 & -- & 4.01 & A0V & Herbig Ae  & 1 \\
HD~223075 & 2346235+032912 & 174.197 $\pm$ 4.056 & 44.205 $\pm$ 0.808 & 0.72 & 0.91 & CII    & Carbon Star  & 5 \\
\hline
\end{tabular}
\end{minipage}
\end{table*}

\subsection{Photospheric SED Fitting and Extinction Determination}

The photospheric flux densities of the $K_{\rm S}-$[18]-selected candidates were estimated 
from the Kurucz model \citep{kurucz92} 
fitted to the 2MASS $JHK_{\rm S}$-band photometry of the stars taking account of the extinction in the NIR. 

We adopt the NIR reddening curve given by \cite{fitzpatrick09}, 
which is a generalization of the analytic formula given by \cite{pei92}. 
For $\lambda \ge \lambda_0$, it has a form of
\begin{eqnarray}
\frac{E(\lambda-V)}{E(B-V)} \equiv R_V \left ( \frac{A_\lambda}{A_V} -1 \right )= \frac{0.349 + 2.087R_V}{1+(\lambda/\lambda_0)^\alpha}-R_V, 
\end{eqnarray}
where $\lambda_0 = 0.507~\micron$. 
Thus, the absolute extinction $A_\lambda/A_V$ is given by 
\begin{eqnarray}
\frac{A_\lambda}{A_V}=  \frac{0.349 + 2.087R_V}{1+(\lambda/\lambda_0)^\alpha} \cdot \frac{1}{R_V}. 
\end{eqnarray}
The ratio of the total to selective extinction, $R_V \equiv A(V)/E(B - V)$, and $\alpha$ are free parameters. 
NIR observations show that these parameters can vary from one sight line to another. 
In this work, we use a curve defined by $\alpha = 2.05$ and $R_V = 3.11$. 
These parameters provide a good representation of the mean curve determined 
by \cite{fitzpatrick09} and also of the extinction data taken from the literature 
for the diffuse interstellar medium in the Milky Way.

In our photospheric SED fitting, the scaling factor of luminosity with varying distance and extinction, 
and the extinction $A_V$ are free parameters, 
while the effective temperature and the surface gravity are inferred from the spectral type. 
The derived photospheric flux densities at 9 and 18$~\micron$ 
for each star are listed in Table~\ref{candidates}.


\begin{table*}
\caption{List of debris disk candidates with $18~\micron$ excess. \label{candidates}}
\centering          
\begin{tabular}{cccccccccccc}
\hline \hline
Name & IRC ID & $F_{{\rm obs},9}$ & $F_{{\rm *},9}$ & $\chi_{9}$ & Excess & $F_{{\rm obs},18}$ & $F_{{\rm *},18}$ & $\chi_{18}$ & Excess \\
 &  & (Jy) & (Jy) &  & at $9~\micron$ & (Jy) & (Jy) &  & at $18~\micron$ \\
\hline                    
HD~3003 & 0032440-630153 & $ 0.586 \pm 0.010 $ & 0.639 &  -1.7 &     & $ 0.238 \pm 0.016 $ & 0.137 &    5.9 & Yes \\ 
HD~9672 & 0134378-154035 & $ 0.368 \pm 0.008 $ & 0.395 &  -2.5 &     & $ 0.199 \pm 0.014 $ & 0.084 &    8.2 & Yes \\ 
HD~15407 & 0230506+553254 & $ 0.960 \pm 0.031 $ & 0.295 &  20.9 & Yes & $ 0.497 \pm 0.021 $ & 0.064 &   20.6 & Yes \\ 
HD~39060 & 0547170-510359 & $ 3.216 \pm 0.009 $ & 2.384 &   1.6 &     & $ 4.799 \pm 0.042 $ & 0.511 &   35.1 & Yes \\ 
HD~39415 & 0554415+443007 & $ 0.244 \pm 0.017 $ & 0.085 &   9.2 & Yes & $ 0.274 \pm 0.015 $ & 0.018 &   17.1 & Yes \\ 
HD~64145 & 0753297+264556 & $ 0.762 \pm 0.005 $ & 0.852 &  -0.8 &     & $ 0.303 \pm 0.041 $ & 0.183 &    2.5 & Marginal \\ 
HD~75809 & 0850454-381440 & $ 0.397 \pm 0.012 $ & 0.399 &  -0.1 &     & $ 0.152 \pm 0.021 $ & 0.088 &    3.0 & Yes \\ 
HD~89125 & 1017143+230621 & $ 0.975 \pm 0.003 $ & 0.988 &  -0.1 &     & $ 0.368 \pm 0.025 $ & 0.214 &    3.1 & Yes \\ 
HD~98800 & 1122052-244639 & $ 1.083 \pm 0.028 $ & 0.357 &  25.0 & Yes & $ 6.153 \pm 0.325 $ & 0.077 &   18.7 & Yes \\ 
HD~105209 & 1206526-593529 & $ 0.392 \pm 0.006 $ & 0.274 &  14.9 & Yes & $ 0.236 \pm 0.013 $ & 0.060 &   13.5 & Yes \\ 
HD~106797 & 1217062-654135 & $ 0.266 \pm 0.017 $ & 0.236 &   1.6 &     & $ 0.148 \pm 0.017 $ & 0.050 &    5.7 & Yes \\ 
HD~109573 & 1236009-395211 & $ 0.360 \pm 0.025 $ & 0.291 &   2.6 & Marginal & $ 1.371 \pm 0.022 $ & 0.062 &   59.4 & Yes \\ 
HD~110058 & 1239461-491156 & $ 0.065 \pm 0.016 $ & 0.055 &   0.6 &     & $ 0.084 \pm 0.006 $ & 0.012 &   12.0 & Yes \\ 
HD~113457 & 1305023-642630 & $ 0.146 \pm 0.010 $ & 0.128 &   1.7 &     & $ 0.107 \pm 0.025 $ & 0.027 &    3.2 & Yes \\ 
HD~113766 & 1306357-460201 & $ 1.308 \pm 0.065 $ & 0.149 &  17.8 & Yes & $ 1.428 \pm 0.050 $ & 0.032 &   27.9 & Yes \\ 
HD~121617 & 1357411-470035 & $ 0.103 \pm 0.008 $ & 0.078 &   3.0 & Yes & $ 0.324 \pm 0.027 $ & 0.017 &   11.4 & Yes \\ 
HD~123356 & 1407340-210437 & $ 1.392 \pm 0.046 $ & 1.133 &   4.7 & Yes & $ 0.567 \pm 0.033 $ & 0.248 &    9.5 & Yes \\ 
HD~145263 & 1610551-253121 & $ 0.236 \pm 0.017 $ & 0.042 &  11.4 & Yes & $ 0.486 \pm 0.013 $ & 0.009 &   36.7 & Yes \\ 
HD~165014 & 1804432-205643 & $ 1.362 \pm 0.043 $ & 0.150 &  28.0 & Yes & $ 0.959 \pm 0.004 $ & 0.033 &  224.1 & Yes \\ 
HD~166191 & 1810303-233401 & $ 1.816 \pm 0.127 $ & 0.101 &  13.5 & Yes & $ 2.466 \pm 0.002 $ & 0.022 & 1133.5 & Yes \\ 
HD~167905 & 1818182-232819 & $ 2.364 \pm 0.007 $ & 0.250 & 231.8 & Yes & $ 1.771 \pm 0.078 $ & 0.055 &   22.0 & Yes \\ 
HD~175726 & 1856371+041553 & $ 0.433 \pm 0.005 $ & 0.419 &   1.2 &     & $ 0.215 \pm 0.034 $ & 0.090 &    3.7 & Yes \\ 
HD~176137 & 1858451+020707 & $ 5.870 \pm 1.112 $ & 1.319 &   4.1 & Yes & $ 4.995 \pm 0.600 $ & 0.299 &    7.8 & Yes \\ 
HD~181296 & 1922512-542526 & $ 0.622 \pm 0.016 $ & 0.626 &  -0.1 &     & $ 0.340 \pm 0.021 $ & 0.134 &    9.3 & Yes \\ 
\hline                  
\end{tabular}
\end{table*}

\subsection{Selected Debris Disk Candidates}

\begin{figure*}
\centering
\includegraphics[width=4.5cm]{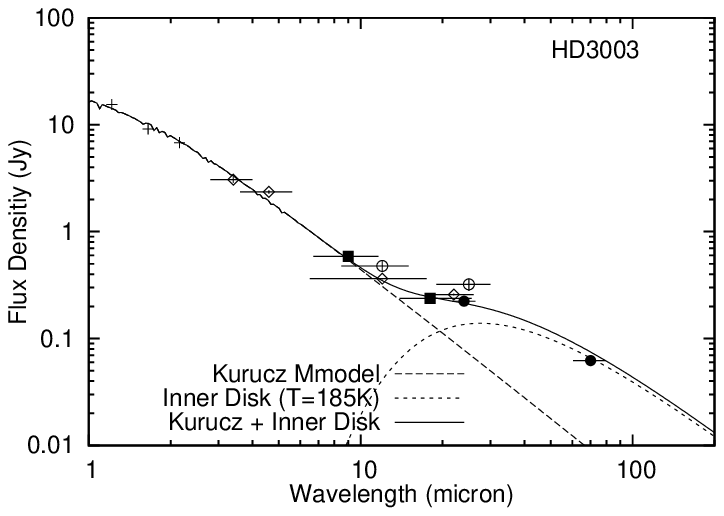}
\includegraphics[width=4.5cm]{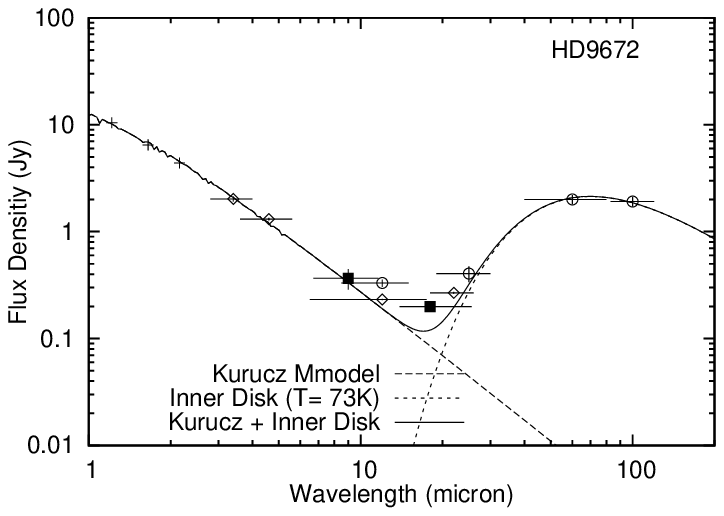}
\includegraphics[width=4.5cm]{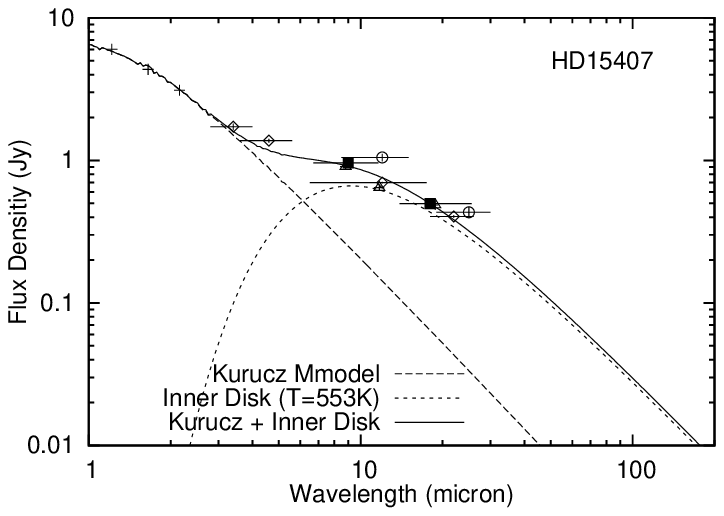}
\includegraphics[width=4.5cm]{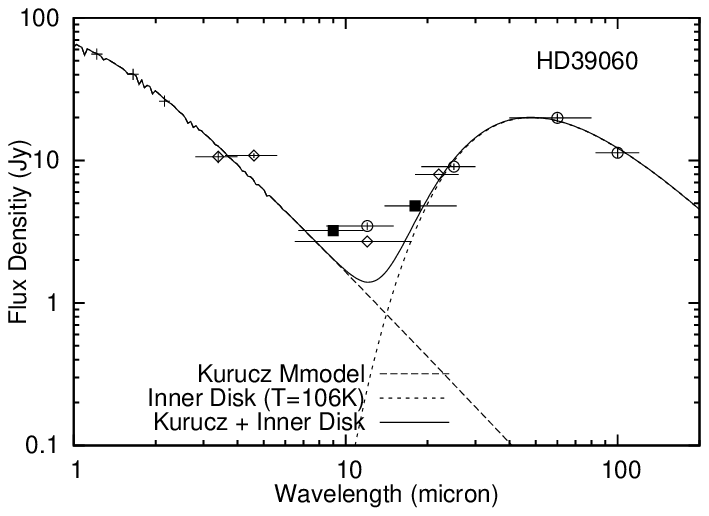}
\includegraphics[width=4.5cm]{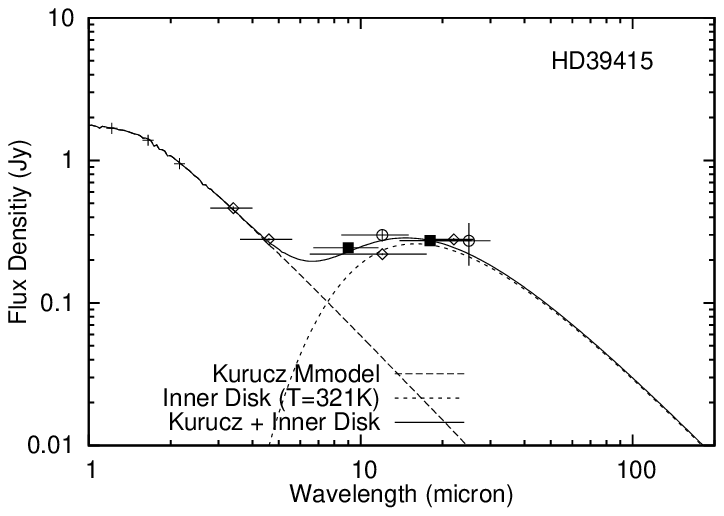}
\includegraphics[width=4.5cm]{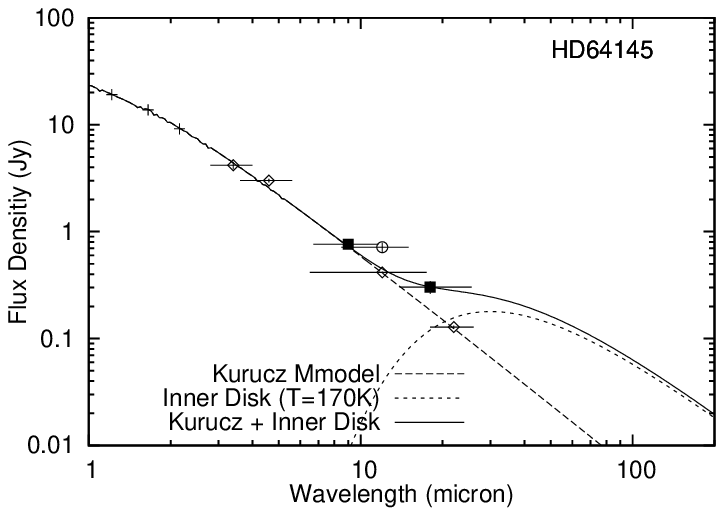}
\includegraphics[width=4.5cm]{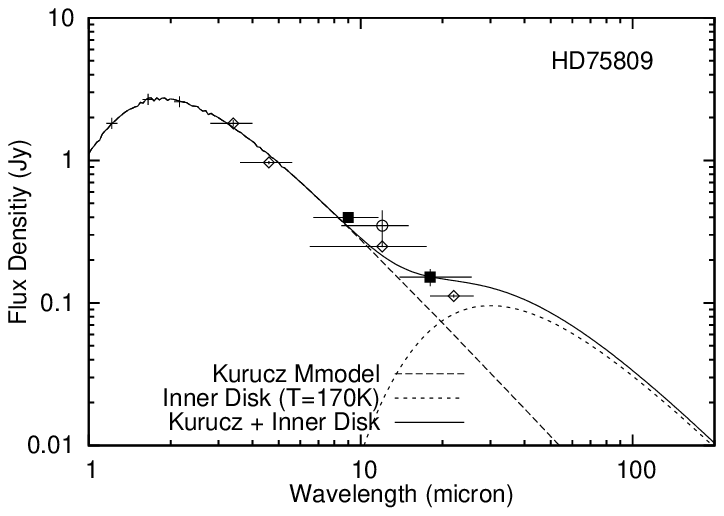}
\includegraphics[width=4.5cm]{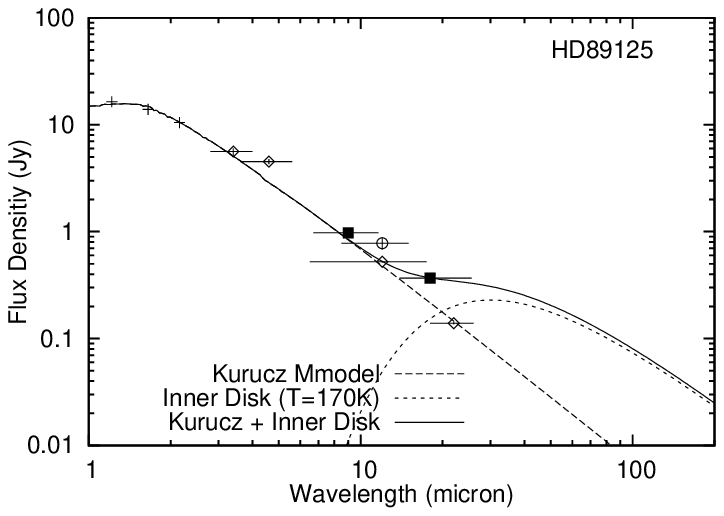}
\includegraphics[width=4.5cm]{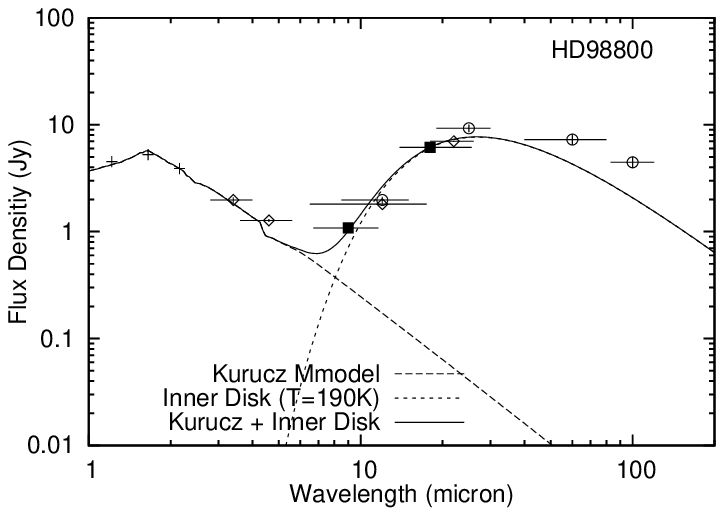}
\includegraphics[width=4.5cm]{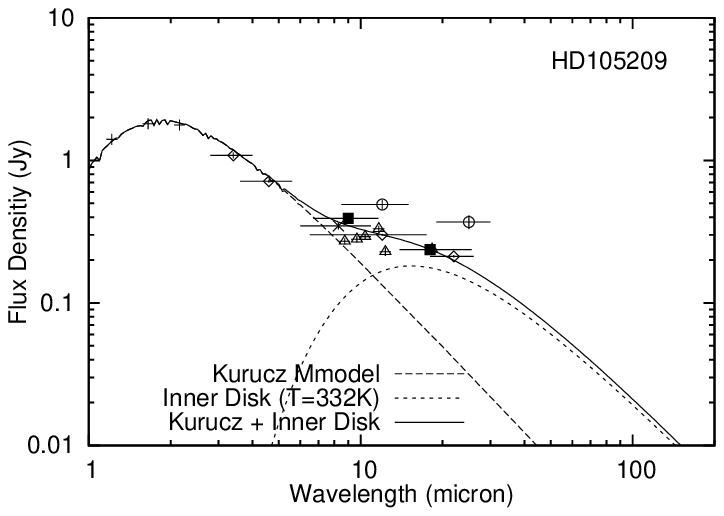}
\includegraphics[width=4.5cm]{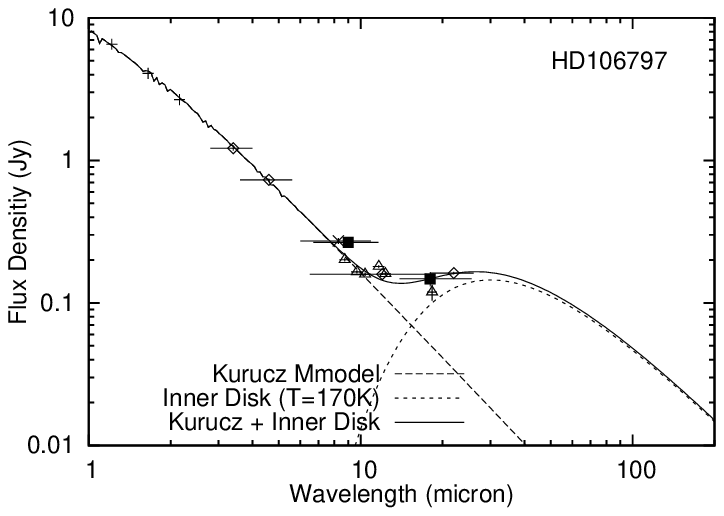}
\includegraphics[width=4.5cm]{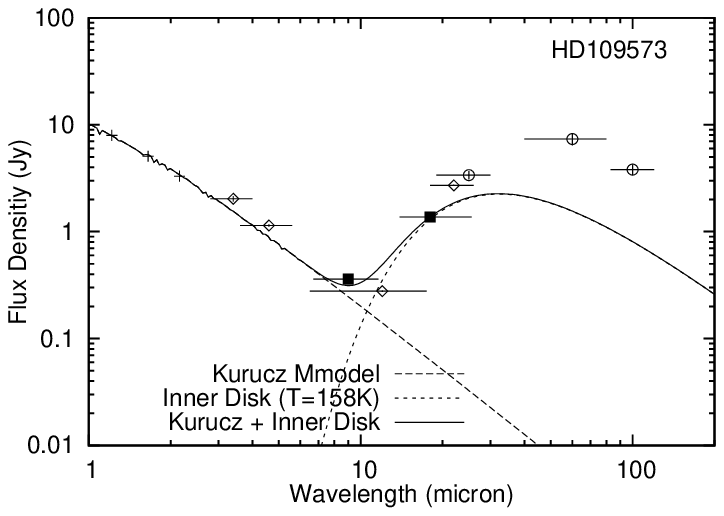}
\includegraphics[width=4.5cm]{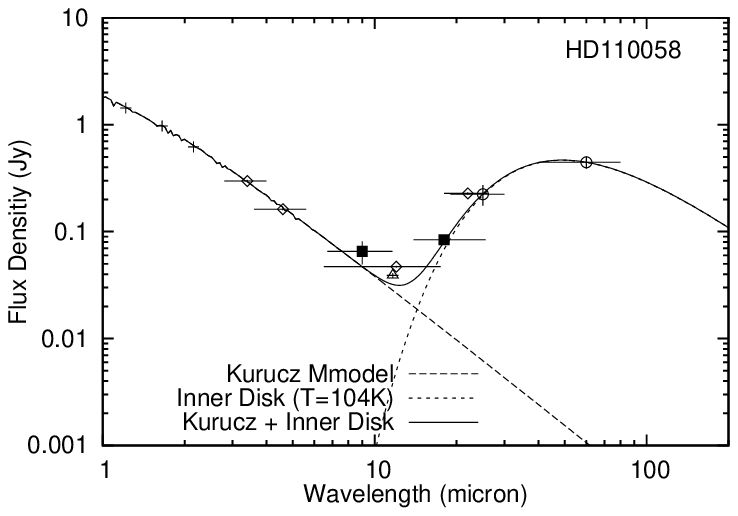}
\includegraphics[width=4.5cm]{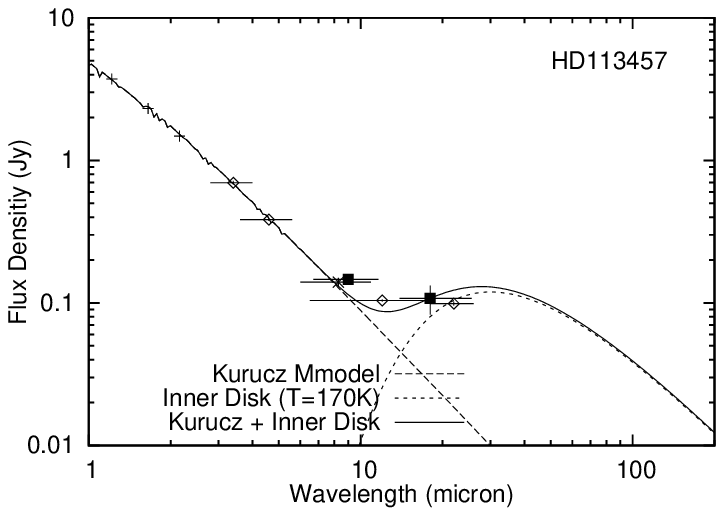}
\includegraphics[width=4.5cm]{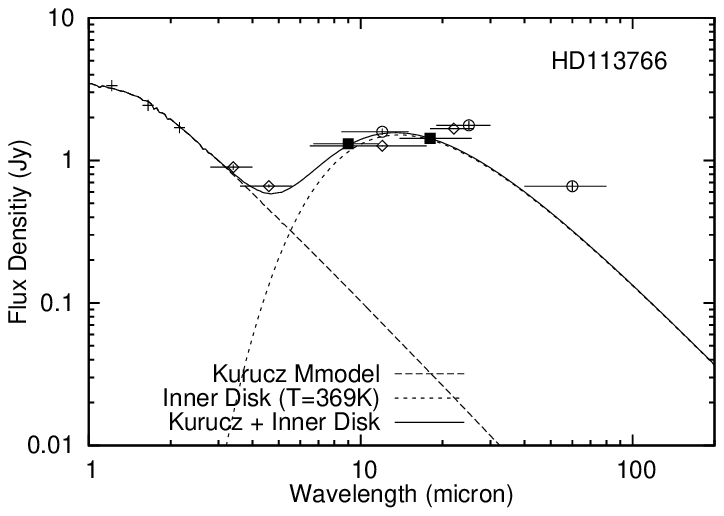}
\includegraphics[width=4.5cm]{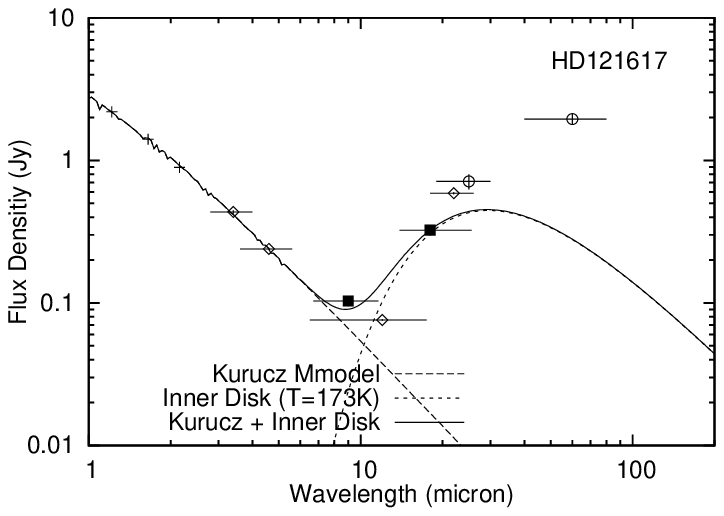}
\includegraphics[width=4.5cm]{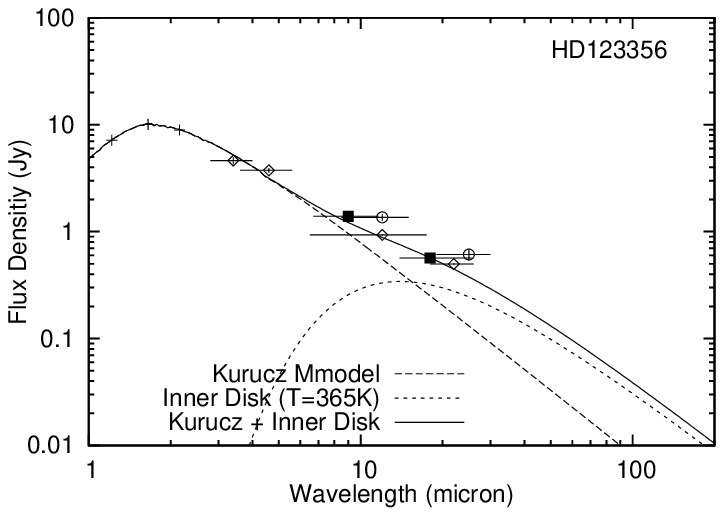}
\includegraphics[width=4.5cm]{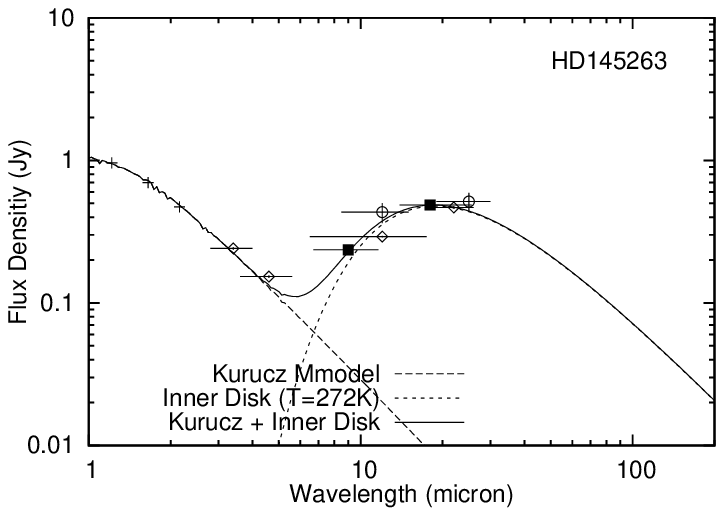}
\includegraphics[width=4.5cm]{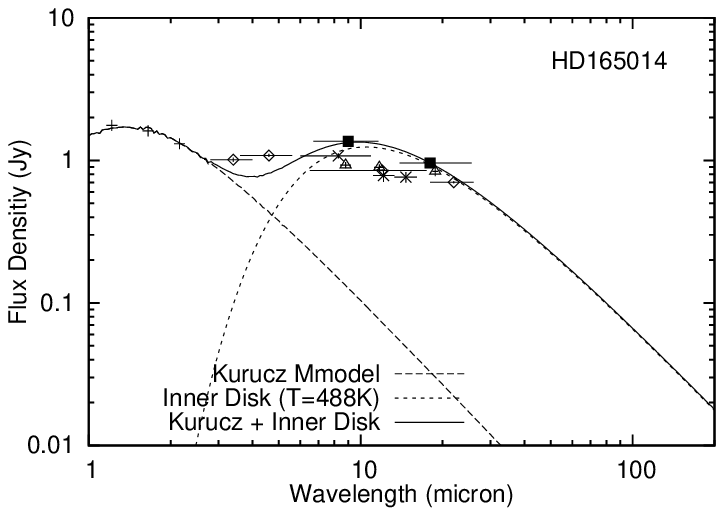}
\includegraphics[width=4.5cm]{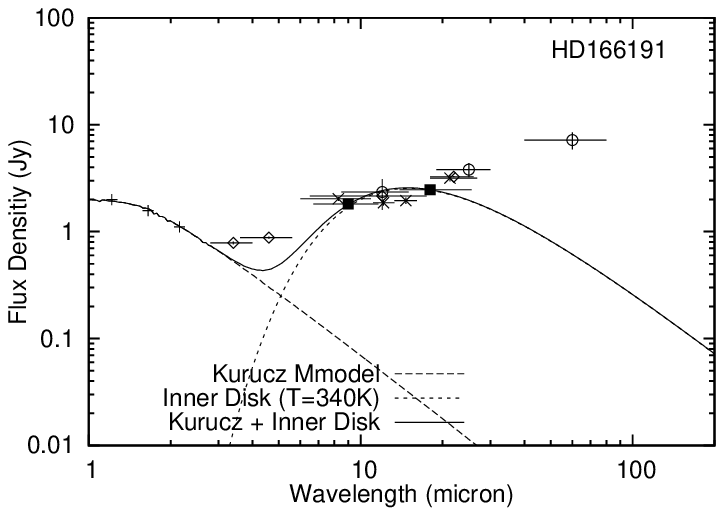}
\includegraphics[width=4.5cm]{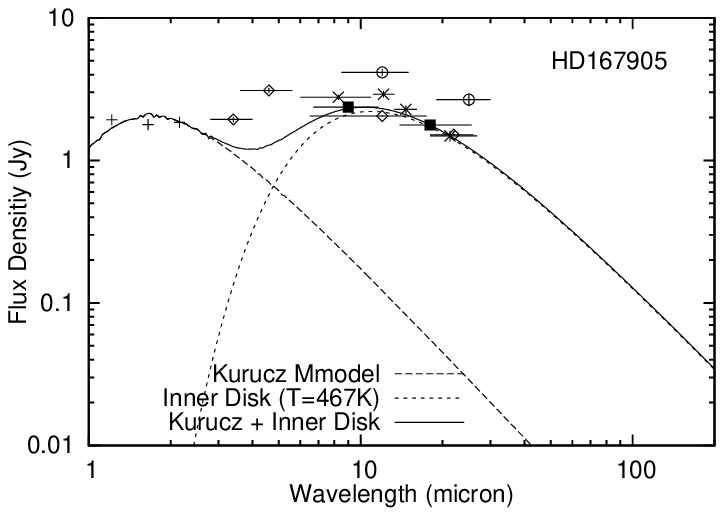}
\includegraphics[width=4.5cm]{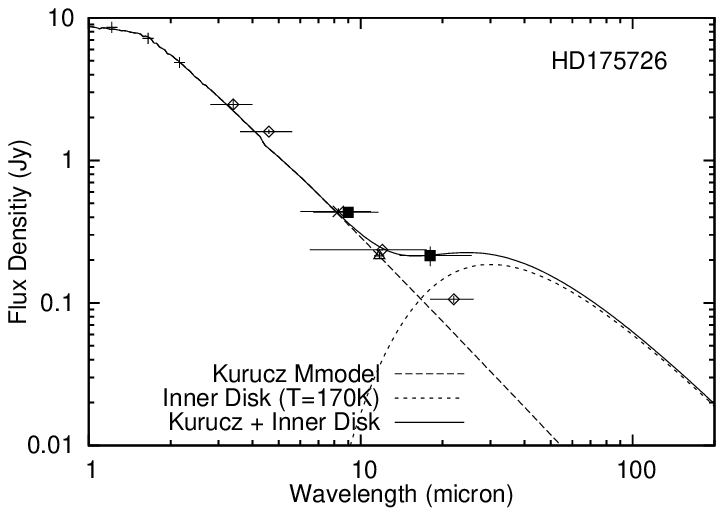}
\includegraphics[width=4.5cm]{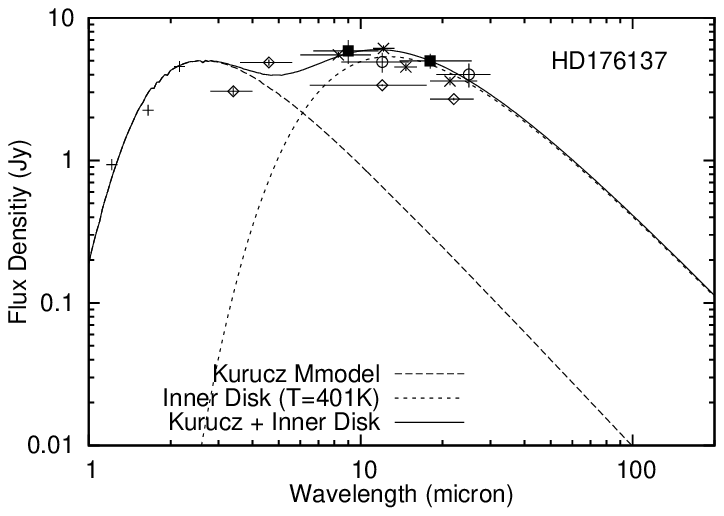}
\includegraphics[width=4.5cm]{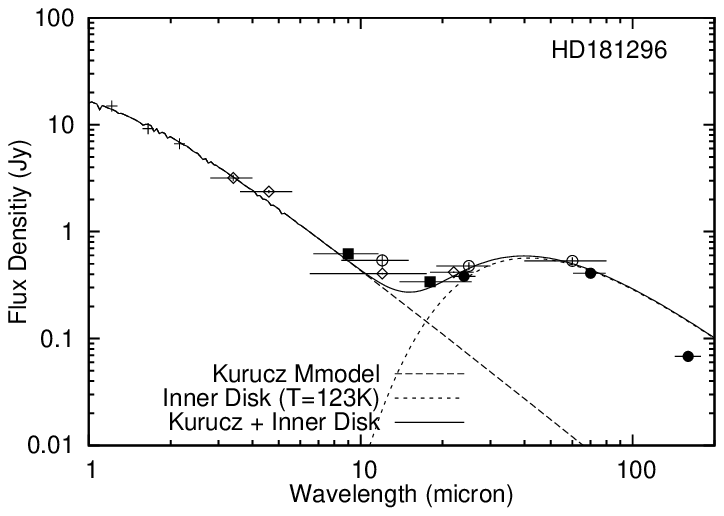}
\caption{NIR--FIR SED of our debris disk candidates with $18~\micron$ excess. 
The pluses, filled squares, open circles, crosses, open diamonds, open triangles, and filled circles indicate 
the photometry with the 2MASS, {\it AKARI}/IRC, {\it IRAS}, {\it MSX}, {\it WISE}, 
ground-based observations (Subaru/COMICS and Gemini/T-ReCS), and {\it Spitzer}/MIPS, repectively. 
We also plot contributions of the photosphere (dashed lines) 
and those of inner disk component (dotted lines) fitted with {\it AKARI}/IRC, {\it IRAS}, and {\it Spitzer}/MIPS measurements 
by assuming that the debris dust emits blackbody emission. 
\label{SEDfit}}
\end{figure*}

The final catalogue contains 24 {\it AKARI}-identified debris disk candidates showing $18~\micron$ excesses. 
Among these, we identify 8 stars (HD~64145, HD~75809, HD~89125, HD~106797, HD~113457, 
HD~165014, HD~175726, and HD~176137) are new debris disk candidates, 
and the excess emission of 2 stars (HD~15407 and HD~39415), 
which have never been confirmed following their
tentative discovery with {\it IRAS} \citep{oudmaijer92}, is confirmed by our {\it AKARI} survey.
The SEDs of the 24 identified debris disk candidates are shown in Figure~\ref{SEDfit}.
We summarise the observed and photospheric flux densities ($F_{\rm obs}$ and $F_*$, respectively) 
and excess significance ($\chi \equiv (F_{\rm obs}-F_*)/\sigma$) at 9 and $18~\micron$ 
of all the candidates in Table~\ref{candidates}, 
where $\sigma$ is the uncertainty of the excess level calculated from the errors of the {\it AKARI}/IRC PSC photometry 
and in the estimation of the photospheric emission from the 2MASS photometry for each source. 
13 stars out of the 24 identified debris disk candidates also show excess at $9~\micron$. 
It should be noted that since the uncertainty in the flux density for a single source is calculated 
as deviation of photometry of multiple detections in the {\it AKARI}/IRC PSC, 
the uncertainty may possibly be too small in case the number of detections is small, 
such as the cases of $18~\micron$ photometry of HD~165014 and HD~166191. 
The median value of the uncertainty at $18~\micron$ in our debris disk candidates is 0.022~Jy. 
Therefore the $18~\micron$ excess towards HD~165014 and HD~166191 will still be significant 
even if we assume the median values as the photometric uncertainties for the two stars. 

We list the stellar parameters (spectral type, stellar mass, luminosity, age, and distance) 
of all the candidates in Table~\ref{stellar}. 
The spectral type of the stars is all taken from the SIMBAD database. 
For HD~3003, HD~9672, HD~15407, HD~39060, HD~89125, HD~98800, HD~106797, HD~109573, HD~110058, HD~145263, and HD~181296, 
we take stellar masses and luminosities from literatures. 
For the other stars whose stellar masses and luminosities are unavailable from the literature, 
we estimate them from their spectral types according to \cite{cox00}. 

\begin{table}
\begin{minipage}[t]{\columnwidth}
\caption{Stellar parameters of debris disk candidates with $18~\micron$ excess. \label{stellar}}
\centering
\renewcommand{\footnoterule}{}  
\begin{tabular}{ccccccc}
\hline \hline
Name & Spectral & $M_*$ & $L_*$ & Age & $d$ & 
Ref.\footnote{References. (1) \cite{wyatt07b}; (2) \cite{zuckerman00}; (3) \cite{fujiwara09}; (4) \cite{dezeeuw99}; 
(5) \cite{cox00}; (6) \cite{crifo97}; (7) \cite{honda04}; (8) \cite{holmberg09}; (9) \cite{melis10}; 
(10) \cite{uzpen07}; (11) \cite{takeda07}; (12) \cite{bruntt09}; (13) \cite{wyatt07}.} \\
 & Type & ($M_\odot$) & ($L_\odot$) & (Myr) & (pc) & \\
\hline
HD~3003 & A0V & 2.9 & 21.0 & 50 & 47 & 1 \\ 
HD~9672 & A1V & 2.7 & 22.8 & 20 & 61 & 1 \\ 
HD~15407 & F3V & 1.4 & 3.9 & 80--2100 & 55 & 8,9 \\ 
HD~39060 & A6V & 1.8 & 8.7 & $12_{-4}^{+8}$ & 19 & 6 \\ 
HD~39415 & F5V & 1.4 & 4.3 & -- & -- & 5 \\ 
HD~64145 & A3V & 2.5 & 39.8 & -- & 78 & 5 \\ 
HD~75809 & F3V & 1.4 & 4.3 & -- & -- & 5 \\ 
HD~89125 & F8Vbw & 0.98 & 2.19 & 6457 & 23 & 11 \\ 
HD~98800 & K5V & 0.67 & 0.7 & 10 & 47 & 13 \\ 
HD~105209 & A1V & 2.9 & 68.7 & -- & -- & 5 \\ 
HD~106797 & A0V & 2.9 & 42.4 & 10--20 & 96 & 3,4 \\ 
HD~109573 & A0V & 2.9 & 24.3 & 8 & 67 & 1 \\ 
HD~110058 & A0V & 2.9 & 10.0 & 10 & 100 & 1 \\ 
HD~113457 & A0V & 2.9 & 68.7 & 10--20 & 95 & 4,5 \\ 
HD~113766 & F4V & 1.4 & 4.3 & 10--20 & 123 & 10 \\ 
HD~121617 & A1V & 2.9 & 68.7 & -- & 120 & 5 \\ 
HD~123356 & G1V & 1.05 & 1.4 & -- & -- & 5 \\ 
HD~145263 & F0V & 1.6 & 4.1 & 8--10 & 116 & 7 \\ 
HD~165014 & F2V & 1.6 & 7.2 & -- & -- & 5 \\ 
HD~166191 & F4V & 1.4 & 4.3 & -- & -- & 5 \\ 
HD~167905 & F3V & 1.4 & 4.3 & -- & 53 & 5 \\ 
HD~175726 & G5 & 1.01 & 1.2 & $4800\pm3500$ & 27 & 12 \\ 
HD~176137 & F5 & 1.4 & 4.3 & -- & -- & 5 \\ 
HD~181296 & A0Vn & 2.9 & 22.0 & $12_{-4}^{+8}$ & 48 & 2 \\ 
\hline
\end{tabular}
\end{minipage}
\end{table}

\section{Additional Photometric Measurements of Debris Disk Candidates}

\subsection{Subaru/COMICS and Gemini/T-ReCS Follow-up Observations}

HD~15407, HD~165014, and HD~175726 were observed with the COoled Mid-Infrared Camera
and Spectrometer \citep[COMICS;][]{kataza00,okamoto02,sako03} 
mounted on the 8 m Subaru Telescope during 2007 June and July. 
Imaging observations at the $8.8~\micron$ ($\Delta\lambda =0.8~\micron$), 
$9.7~\micron$ ($\Delta\lambda =0.9~\micron$), 
$10.5~\micron$ ($\Delta\lambda =1.0~\micron$), $11.7~\micron$ ($\Delta\lambda =1.0~\micron$), 
$12.4~\micron$ ($\Delta\lambda =1.2~\micron$), and $18.8~\micron$ ($\Delta\lambda =0.9~\micron$) 
bands were carried out. The pixel scale was $0\farcs13$ pixel$^{-1}$. 
To cancel out the background radiation, a secondary mirror chopping method was used. 
We used standard stars from \citet{cohen99} as a flux calibrator, 
and determined the reference point-spread functions (PSFs) from the observations.
We observed standard stars before or after the observations of the target star in the same manner as the target stars. 
The parameters of the COMICS observations are summarised in Table~\ref{comicsobs}.

HD~106797, HD~105209, and HD~110058 were observed 
with the Thermal-Region Camera Spectrograph \citep[T-ReCS;][]{telesco98}, 
mounted on the 8 m Gemini South Telescope during 2007 June and July. 
Imaging observations at the $8.8~\micron$ ($\Delta\lambda =0.8~\micron$; Si2), 
$9.7~\micron$ ($\Delta\lambda =0.9~\micron$; Si3), 
$10.4~\micron$ ($\Delta\lambda =1.0~\micron$; Si4), $11.7~\micron$ ($\Delta\lambda =1.1~\micron$; Si5), 
$12.3~\micron$ ($\Delta\lambda =1.2~\micron$; Si6), and $18.3~\micron$ ($\Delta\lambda =1.5~\micron$; Qa) 
bands were carried out. The pixel scale was $0\farcs09$ pixel$^{-1}$. 
Procedures for the observations were basically the same as for the COMICS observations. 
The parameters of the T-ReCS observations are summarised in Table~\ref{trecsobs}.

For the data reduction, we used our own reduction tools and 
IRAF\footnote{IRAF is distributed by the National Optical Astronomy Observatory, which is operated 
by the Association of Universities for Research in Astronomy (AURA) under cooperative agreement 
with the National Science Foundation.}. 
The standard chop-nod pair subtraction and the shift-and-add method 
in units of 0.1 pixel were employed. 
We applied an air mass correction by estimating the difference between the target and standard stars 
in atmospheric absorption using ATRAN \citep{lord92}.
The difference in the air masses between the target and standard stars was small ($\lesssim 0.1$) in general 
and the correction factor in each band was less than $5$~\%.

The derived flux densities are summarised in Tables~\ref{comics} and \ref{trecs} 
for the COMICS and T-ReCS targets, respectively. 
All of them are in good agreement with the flux densities of the {\it AKARI} observations 
taking account of the narrow band widths of filters of the ground-based observations 
and possible dust features in these spectral ranges \citep{fujiwara09,fujiwara10,fujiwara12a}.


\begin{table}
\caption{Summary of Subaru/COMICS photometry. \label{comics}}
\centering                          
\begin{tabular}{ccccc}        
\hline\hline                 
Object    & N8.8      & N11.7 & Q18.8 \\
          & (mJy)     & (mJy) & (mJy) \\
\hline                        
HD~15407  & $904 \pm  90 $ & $644 \pm  64 $ & $486 \pm  73 $ \\
HD~165014 & $927 \pm  93 $ & $894 \pm  89 $ & $838 \pm 126 $ \\
HD~175726 & --            & $214 \pm  20 $ & --            \\
\hline                                   
\end{tabular}
\end{table}


\begin{table*}
\caption{Summary of Gemini/T-ReCS photometry. \label{trecs}}
\centering                          
\begin{tabular}{ccccccc}        
\hline\hline                 
Object    & Si2 8.8   & Si3 9.7  & Si4 10.4 & Si5 11.7  & Si6 12.3  & Qa 18.8  \\
          & (mJy)     & (mJy)    & (mJy)    & (mJy)     & (mJy)     & (mJy)    \\
\hline                        
HD~105209 & $271 \pm 27$ & $280 \pm 28$ & $292 \pm 29$ & $331 \pm 33$ & $299 \pm 30$ & $238 \pm 31$ \\
HD~106797 & $201 \pm 20$ & $165 \pm 17$ & $159 \pm 15$ & $180 \pm 14$ & $160 \pm 13$ & $119 \pm 59$ \\
HD~110058 & --           & --           & --           & $39 \pm 4$   & --           & --           \\
\hline                                   
\end{tabular}
\end{table*}

\subsection{{\it IRAS}, {\it MSX}, {\it Spitzer} and {\it WISE} Photometry}

We searched for counterparts whose positional offsets were smaller than $1\arcmin$ 
in {\it IRAS} \citep{beichman88} Point Source Catalogue (PSC) and Faint Source Catalogue (FSC), 
and MSX6C Infrared Point Source Catalogue \citep{eagan03} of the Midcourse Space Experiment 
\citep[{\it MSX}{\rm ;}][]{price01} via VizieR for each debris disk candidate. 
The results are shown in Tables~\ref{IRAS} and \ref{MSX} for {\it IRAS} and {\it MSX}, respectively. 
We also searched for {\it Spitzer}/MIPS data from published literatures 
and found those of HD~3003 \citep{smith06} and HD~181269 \citep{su06,rebull08}, as shown in Table~\ref{MIPS}. 

Recent all-sky observations in the MIR by {\it Wide-field Infrared Survey Explorer} ({\it WISE}) 
have provided the All-Sky Source Catalogue \citep{cutri12}, which is similar to {\it AKARI}/IRC PSC 
and is worth comparing. 
We searched for counterparts of our MIR excess stars whose positional offsets were smaller than $5\arcsec$ 
in the {\it WISE} All-Sky Source Catalogue for comparison. 
The results are shown in Table~\ref{WISE}. 
All of our debris disk candidates with $18~\micron$ excesses have a counterpart 
in the the {\it WISE} All-Sky Source Catalog.


\begin{table*}
\caption{IRAS photometry of the debris disk candidates with $18~\micron$ excess. \label{IRAS}}
\centering                          
\begin{tabular}{ccccccc}        
\hline\hline                 
Name & IRAS12 & IRAS25 & IRAS60 & IRAS100 & {\it IRAS} ID & Catalogue \\
     & (Jy)   & (Jy)   & (Jy)   & (Jy)    &         &         \\
\hline                        
HD~3003 & $  0.48 \pm 0.06 $ & $  0.32 \pm 0.04 $ & $ <0.40          $ & $ <1.00              $ & 00304-6318 &        PSC \\
HD~9672 & $  0.33 \pm 0.03 $ & $  4.06 \pm 0.07 $ & $  2.00 \pm 0.20 $ & $  1.91 \pm 0.19     $ & 01321-1555 &        PSC \\
HD~15407 & $  1.05 \pm 0.06 $ & $  0.43 \pm 0.04 $ & $ <0.40          $ & $ <1.14              $ & 02272+5519 &        PSC \\
HD~15407 & $  1.05 \pm 0.06 $ & $  0.43 \pm 0.04 $ & $ <0.40          $ & $ <1.14              $ & 02272+5519 &        PSC \\
HD~39060 & $  3.46 \pm 0.21 $ & $  9.05 \pm 0.45 $ & $  19.9 \pm 1.4  $ & $  11.3 \pm 0.90     $ & 05460-5104 &        PSC \\
HD~39415 & $  0.30 \pm 0.03 $ & $  0.27 \pm 0.09 $ & $ <0.40          $ & $ <1.23              $ & 05510+4429 &        PSC \\
HD~64145 & $  0.72 \pm 0.07 $ & $ <0.64          $ & $ <0.40          $ & $ <1.03              $ & 07504+2653 &        PSC \\
HD~75809 & $  0.35 \pm 0.10 $ & $ <0.25          $ & $ <0.41          $ & $ <5.28              $ & 08488-3803 &        PSC \\
HD~89125 & $  0.78 \pm 0.10 $ & $ <0.25          $ & $ <0.40          $ & $ <1.00              $ & 10144+2321 &        PSC \\
HD~98800 & $  1.98 \pm 0.14 $ & $  9.28 \pm 0.74 $ & $  7.28 \pm 0.80 $ & $  4.46 \pm 0.54     $ & 11195-2430 &        PSC \\
HD~105209 & $  0.49 \pm 0.04 $ & $  0.37 \pm 0.40 $ & $ <1.78          $ & $ <25.0              $ & 12043-5919 &        PSC \\
HD~109573 & $ <0.43          $ & $  3.38 \pm 0.30 $ & $  7.36 \pm 0.96 $ & $  3.81 \pm 0.42     $ & 12333-3935 &        PSC \\
HD~110058 & $ <0.38          $ & $  0.22 \pm 0.05 $ & $  0.45 \pm 0.06 $ & $ <1.04              $ & 12369-4855 &        PSC \\
HD~113766 & $  1.59 \pm 0.08 $ & $  1.76 \pm 0.12 $ & $  0.66 \pm 0.05 $ & $ <1.00              $ & 13037-4545 &        PSC \\
HD~121617 & $ <0.25          $ & $  0.72 \pm 0.09 $ & $  1.95 \pm 0.18 $ & $ <1.45              $ & 13545-4645 &        PSC \\
HD~123356 & $  1.36 \pm 0.10 $ & $  0.61 \pm 0.07 $ & $ <0.40          $ & $ <1.00              $ & 14047-2050 &        PSC \\
HD~145263 & $  0.43 \pm 0.07 $ & $  0.52 \pm 0.08 $ & $ <1.62          $ & $ <3.49              $ & 16078-2523 &        FSC \\
HD~166191 & $  2.35 \pm 0.75 $ & $  3.80 \pm 0.57 $ & $  7.18 \pm 1.29 $ & $ <208               $ & 18074-2334 &        PSC \\
HD~167905 & $  4.14 \pm 0.25 $ & $  2.67 \pm 0.27 $ & $ <2.76          $ & $ <59.2              $ & 18152-2329 &        PSC \\
HD~176137 & $  4.91 \pm 0.79 $ & $  4.01 \pm 0.76 $ & $ <11.6          $ & $ <127               $ & 18562+0202 &        PSC \\
HD~181296 & $  0.54 \pm 0.08 $ & $  0.48 \pm 0.04 $ & $  0.53 \pm 0.04 $ & $ <1.00              $ & 19188-5431 &        PSC \\
\hline                                   
\end{tabular}
\end{table*}


\begin{table*}
\caption{MSX photometry of the debris disk candidates with $18~\micron$ excess. \label{MSX}}
\centering                          
\begin{tabular}{cccccc}        
\hline\hline                 
Name & Band A ($8.28~\micron$) & Band C ($12.13~\micron$) & Band D ($14.65~\micron$) & Band E ($21.3~\micron$) & {\it MSX} ID \\
     & (Jy)   & (Jy)   & (Jy)   & (Jy)    &     \\
\hline                        
HD~105209 & $  0.347 \pm 0.018 $ & $ <0.952              $ & $ <0.728           $ & $ <2.082           $ & G297.3128+02.7932 \\
HD~106797 & $  0.272 \pm 0.013 $ & $ <0.449              $ & $ <0.354           $ & $ <1.039           $ & G299.4048-03.0570 \\
HD~113457 & $  0.140 \pm 0.010 $ & $ <0.601              $ & $ <0.466           $ & $ <1.349           $ & G304.3984-01.6086 \\
HD~165014 & $  1.077 \pm 0.045 $ & $  0.783 \pm 0.074    $ & $  0.764 \pm 0.065 $ &          --          & G009.0807+00.3009 \\
HD~166191 & $  2.028 \pm 0.081 $ & $  1.863 \pm 0.117    $ & $  1.953 \pm 0.129 $ & $  3.166 \pm 0.212 $ & G007.4400-02.1430 \\
HD~167905 & $  2.772 \pm 0.114 $ & $  2.912 \pm 0.163    $ & $  2.279 \pm 0.148 $ & $  1.487 \pm 0.126 $ & G008.3752-03.6697 \\
HD~175726 & $  0.439 \pm 0.020 $ & $ <0.573              $ & $ <0.450           $ & $ <1.262           $ & G037.3180+00.7938 \\
HD~176137 & $  5.497 \pm 0.236 $ & $  6.102 \pm 0.305    $ & $  4.517 \pm 0.276 $ & $  3.601 \pm 0.230 $ & G035.6512-00.6602 \\
\hline                                   
\end{tabular}
\end{table*}


\begin{table}
\begin{minipage}[t]{\columnwidth}
\caption{Spitzer/MIPS photometry of the debris disk candidates with $18~\micron$ excess. \label{MIPS}}
\centering
\renewcommand{\footnoterule}{}  
\begin{tabular}{ccccc}
\hline \hline
Name & MIPS24 & MIPS70 & MIPS160 & 
Ref.\footnote{References. (1) \cite{smith06}; (2) \cite{su06}; (3) \cite{rebull08}.} \\
 & (mJy) & (mJy) & (mJy) & \\
\hline
    HD~3003 & $ 224 \pm 9 $ & $ 62  \pm 5  $ &  --  & 1           \\
  HD~181296 & $ 382 \pm 7 $ & $ 409 \pm 42 $ & $ 68 $ & 2,3 \\
\hline
\end{tabular}
\end{minipage}
\end{table}


\begin{table*}
\caption{WISE photometry of the debris disk candidates with $18~\micron$ excess. \label{WISE}}
\centering                          
\begin{tabular}{cccccc}        
\hline\hline                 
Name & W1 ($3.4~\micron$) & W2 ($4.6~\micron$) & W3 ($12~\micron$) & W4 ($22~\micron$) & {\it WISE} ID \\
     & (Jy)   & (Jy)   & (Jy)   & (Jy)    &     \\
\hline                        
HD~3003 & $  3.066 \pm  0.198 $ & $  2.356 \pm  0.084 $ & $  0.363 \pm  0.005 $ & $  0.258 \pm  0.005 $ &  J003244.02-630154.0 \\ 
HD~9672 & $  2.026 \pm  0.115 $ & $  1.311 \pm  0.033 $ & $  0.232 \pm  0.003 $ & $  0.268 \pm  0.006 $ &  J013437.83-154034.8 \\ 
HD~15407 & $  1.721 \pm  0.093 $ & $  1.376 \pm  0.040 $ & $  0.699 \pm  0.010 $ & $  0.404 \pm  0.009 $ &  J023050.75+553253.3 \\ 
HD~39060 & $ 10.603 \pm  1.023 $ & $ 10.808 \pm  0.406 $ & $  2.698 \pm  0.017 $ & $  7.972 \pm  0.081 $ &  J054717.01-510357.5 \\ 
HD~39415 & $  0.463 \pm  0.013 $ & $  0.279 \pm  0.005 $ & $  0.220 \pm  0.003 $ & $  0.279 \pm  0.006 $ &  J055441.52+443007.5 \\ 
HD~64145 & $  4.182 \pm  0.328 $ & $  3.010 \pm  0.113 $ & $  0.416 \pm  0.006 $ & $  0.128 \pm  0.004 $ &  J075329.80+264556.5 \\ 
HD~75809 & $  1.819 \pm  0.096 $ & $  0.966 \pm  0.023 $ & $  0.249 \pm  0.003 $ & $  0.112 \pm  0.003 $ &  J085045.44-381439.9 \\ 
HD~89125 & $  5.611 \pm  0.429 $ & $  4.530 \pm  0.235 $ & $  0.522 \pm  0.007 $ & $  0.139 \pm  0.003 $ &  J101714.20+230621.6 \\ 
HD~98800 & $  1.976 \pm  0.116 $ & $  1.273 \pm  0.038 $ & $  1.807 \pm  0.020 $ & $  6.995 \pm  0.065 $ &  J112205.23-244639.4 \\ 
HD~105209 & $  1.086 \pm  0.051 $ & $  0.715 \pm  0.016 $ & $  0.301 \pm  0.004 $ & $  0.212 \pm  0.004 $ &  J120652.71-593529.7 \\ 
HD~106797 & $  1.217 \pm  0.048 $ & $  0.731 \pm  0.014 $ & $  0.159 \pm  0.002 $ & $  0.162 \pm  0.003 $ &  J121706.26-654134.7 \\ 
HD~109573 & $  2.028 \pm  0.119 $ & $  1.143 \pm  0.029 $ & $  0.278 \pm  0.004 $ & $  2.714 \pm  0.035 $ &  J123600.95-395210.8 \\ 
HD~110058 & $  0.298 \pm  0.007 $ & $  0.162 \pm  0.003 $ & $  0.047 \pm  0.001 $ & $  0.228 \pm  0.005 $ &  J123946.17-491155.6 \\ 
HD~113457 & $  0.697 \pm  0.021 $ & $  0.384 \pm  0.008 $ & $  0.104 \pm  0.002 $ & $  0.099 \pm  0.003 $ &  J130501.99-642629.8 \\ 
HD~113766 & $  0.896 \pm  0.039 $ & $  0.660 \pm  0.015 $ & $  1.265 \pm  0.011 $ & $  1.672 \pm  0.015 $ &  J130635.77-460202.0 \\ 
HD~121617 & $  0.435 \pm  0.013 $ & $  0.239 \pm  0.004 $ & $  0.076 \pm  0.001 $ & $  0.588 \pm  0.010 $ &  J135741.10-470034.4 \\ 
HD~123356 & $  4.624 \pm  0.358 $ & $  3.747 \pm  0.169 $ & $  0.932 \pm  0.012 $ & $  0.497 \pm  0.009 $ &  J140734.01-210437.5 \\ 
HD~145263 & $  0.241 \pm  0.005 $ & $  0.153 \pm  0.003 $ & $  0.292 \pm  0.004 $ & $  0.468 \pm  0.009 $ &  J161055.09-253121.9 \\ 
HD~165014 & $  1.010 \pm  0.036 $ & $  1.083 \pm  0.026 $ & $  0.851 \pm  0.011 $ & $  0.703 \pm  0.012 $ &  J180443.14-205644.6 \\ 
HD~166191 & $  0.787 \pm  0.026 $ & $  0.879 \pm  0.020 $ & $  2.157 \pm  0.028 $ & $  3.271 \pm  0.046 $ &  J181030.32-233400.6 \\ 
HD~167905 & $  1.938 \pm  0.086 $ & $  3.100 \pm  0.125 $ & $  2.052 \pm  0.025 $ & $  1.516 \pm  0.027 $ &  J181818.22-232819.7 \\ 
HD~175726 & $  2.467 \pm  0.176 $ & $  1.593 \pm  0.055 $ & $  0.236 \pm  0.003 $ & $  0.106 \pm  0.006 $ &  J185637.17+041553.6 \\ 
HD~176137 & $  3.055 \pm  0.203 $ & $  4.868 \pm  0.239 $ & $  3.369 \pm  0.044 $ & $  2.691 \pm  0.050 $ &  J185845.10+020706.9 \\ 
HD~181296 & $  3.184 \pm  0.209 $ & $  2.369 \pm  0.078 $ & $  0.405 \pm  0.006 $ & $  0.418 \pm  0.007 $ &  J192251.21-542527.0 \\ 
\hline                                   
\end{tabular}
\end{table*}

\section{Results and Discussion}

\subsection{Debris Disk Frequency}

The derived apparent debris disk frequency (the fraction of the debris disk candidates 
in those stars detected at $18~\micron$ in the {\it AKARI}/IRC All-Sky Survey) 
in our sample is summarised in Table~\ref{frequency}.
We find an overall apparent debris disk frequency of $2.8 \pm 0.6$\% (24/856).
The apparent debris disk frequency for A, F, G, K, and M stars is estimated to be 
$5.6 \pm 1.6$\% (11/196), $3.1 \pm 1.0$\% (10/324), $1.2 \pm 0.8$\% (2/173), $0.7 \pm 0.7$\% (1/144), 0.0\% (0/19), respectively.  
Since it is limited by the sensitivity of the {\it AKARI}/IRC All-Sky Survey, 
the sample is biased towards stars exhibiting MIR excess. 
To estimate the debris fraction properly, 
we count only the excess sample whose photosphere 
would be detectable at $18~\micron$ with the {\it AKARI}/IRC All-Sky Survey 
(i.e. $F_{{\rm *},18} \gtrsim 90$~mJy; HD~3003, HD~39060, HD~64145, HD~75809, HD~89125, HD~123356, HD~175726, HD~176137, and HD~181296) 
among $18~\micron$-detected sources with detectable photosphere, 
and calculate the ``real'' debris disk frequency relative to those with $F_{{\rm *},18} \gtrsim 90$~mJy.
We find an overall debris disk frequency of $1.1 \pm 0.4$\% (9/856).
The debris disk frequency for A, F, G, K, and M stars is derived 
as $2.2 \pm 1.1$\% (4/178), $1.0 \pm 0.6$\% (3/311), $1.2 \pm 0.8$\% (2/169), 0.0\% (0/144), 0.0\% (0/19), respectively, 
suggesting a possible tendency of an increase in the debris disk frequency towards earlier type stars. 
The statistics of the debris disk frequency is summarises in Table~\ref{frequency}.

The excess rates at $24~\micron$ by {\it Spitzer}/MIPS observations for A stars 
and solar-type FGK stars are 32\% (52/160) \citep{su06} 
and 6\% (5/82) \citep{beichman06}, respectively. 
Therefore the debris disk frequency derived in this work is much smaller than 
those estimated by {\it Spitzer}/MIPS observations both for A and FGK stars. 
It should be noted that the uncertainty in {\it Spitzer}/MIPS pointed observations 
is smaller than the present catalogue of the {\it AKARI}/IRC All-Sky Survey 
and thus {\it Spitzer}/MIPS observations detect fainter excess than this work. 
Our thresholds for $18~\micron$ excess detection for the {\it AKARI}/IRC sample in this study are shown in Table~\ref{fitparam}, 
which can potentially detect a debris disk star with $18~\micron$ excess of $F_{\rm obs,18}/F_{\rm *,18} \gtrsim 1.4$.
If we take a debris disk sample with $24~\micron$ excess from the {\it Spitzer}/MIPS sample 
with the same threshold ($F_{\rm obs,24}/F_{\rm *,24} \gtrsim 1.4$) as set for the {\it AKARI}/IRC sample in this work, 
the frequency becomes 14\% (22/160) and 1\% (1/82) for A and FGK stars, respectively. 
The debris disk frequency of FGK stars in this work is derived as $0.8\pm0.3$\% (5/624), 
being consistent with the {\it Spitzer}/MIPS work \citep{beichman06}. 
On the other hand, the frequency of A stars in this work is still smaller than 
that of the {\it Spitzer}/MIPS work \citep{su06}
even after considering the difference in the excess detection threshold. 
The difference might come from a selection effect of the input stars; 
we select only dwarf A stars from the Tycho-2 Spectral Type Catalogue, 
while the \cite{su06} sample includes a number of very young star (age$<$10~Myr) and late B stars, 
which show excess more frequently. 

The frequency of debris disks around M stars in this work is derived as 0.0\% (0/19). 
Although the only known M-type debris disk AU Mic is detected with IRC 
with $1.08 \pm 0.01$ and $0.25 \pm 0.02$~Jy at 9 and $18~\micron$, respectively, 
no MIR excess is found towards the star. 
The non-detection of excess at 9 and $18~\micron$ around AU Mic is consistent 
with MIR observations with {\it Spitzer} \cite[e.g.][]{chen05}.

\subsection{Characteristics of Selected Debris Disk Candidates}

\subsubsection{MIR colour, Inner Temperature, and Inner Radius}

Estimate of the dust temperature is important to infer the radial structures of debris disks.
MIR observations are particularly useful to trace the temperature of the inner hottest region of debris disks.
We derived the dust temperature from the 9 to $18~\micron$ colours of the observed excess emission for each star 
by assuming that the debris dust emits blackbody emission. 
This temperature is supposed to correspond to the dust temperature ($T_{\rm in}$) at the inner radius ($R_{\rm in}$) of the debris disk. 

In the case where excess emission at $9~\micron$ was not detected with {\it AKARI}/IRC, 
the dust temperature should be lower than the temperature for those which show excess at $9~\micron$.  
For the sources whose $9~\micron$ excess was not detected with {\it AKARI}/IRC and 
excess at the wavelengths of $\lambda \gtrsim 24~\micron$ was reported by {\it IRAS} or {\it Spitzer} observations, 
we derived the dust temperature from blackbody fitting to the observed excess emission at $\lambda \gtrsim 24~\micron$. 
In the case that excesses at $9~\micron$ and at $\gtrsim 24~\micron$ were not detected with {\it AKARI}/IRC, {\it IRAS}, and {\it Spitzer}, 
we set an upper limit of $T_{\rm in}$ at 170~K, which is the highest possible temperature of the coldest debris disk 
with 9 and $18~\micron$ excess in the sample ($T_{\rm in} = 158 ^{+11} _{-13}$ for HD~109573). 
We also derived the inner radius of debris disk ($R_{\rm in}$) as 
\begin{eqnarray}
R_{\rm in} = \sqrt{\frac{L_*}{16\pi\sigma_{\rm SB}}}T_{\rm in}^{-2},
\end{eqnarray}
where $L_*$ and $\sigma_{\rm SB}$ is the stellar luminosity and the Stefan-Boltzmann constant, respectively. 
The derived MIR colour, $T_{\rm in}$, and $R_{\rm in}$ are listed in Table~\ref{inner}. 

The inner temperature $T_{\rm in}$ and inner radius $R_{\rm in}$ of A stars range 
from $\lesssim 100$~K to $\sim 330$~K and from $\sim 5$~AU to $> 20$~AU, respectively. 
The median value of the inner temperature among A stars is $\lesssim 170$~K. 
On the other hand, $T_{\rm in}$ and $R_{\rm in}$ of solar-type FGK stars ranges 
from $\lesssim 170$~K to $\sim 500$~K and from $< 1$~AU to $\gtrsim 5$~AU, respectively. 
The median value of the inner temperature among FGK stars is $\sim 350$~K, much higher than that of A stars. 
This suggests that the origin of the observed MIR excess is different 
between A and FGK stars: while the MIR excess emission of A stars in our sample is 
typically contributed from the Wien-side tail of the emission from the cool debris material, 
that of FGK stars originates in the emission from the inner warm debris material.


\begin{table*}
\caption{Inner temperature and radius of debris disk candidates with $18~\micron$ excess. \label{inner}}
\centering                          
\begin{tabular}{cccccccc}        
\hline\hline                 
Name & $F_{{\rm exc},9}$ & $F_{{\rm exc},18}$ & $F_{{\rm exc},9}/F_{{\rm exc},18}$ & $T_{\rm in}$ & $R_{\rm in}$ & $L_{\rm dust}/L_*$ & Cooler \\
     & (Jy)              & (Jy)               &                                    & (K)          & (AU)         &                  & Material \\
\hline                        
HD~3003 & --      & 0.101 & --                  & $ 185 ^{+21} _{-21} $ & $ 10.4 ^{+ 2.8} _{- 2.0} $  & $7 \times 10^{-5}$    &  \\ 
HD~9672 & --      & 0.115 & --                  & $  73 ^{+ 4} _{- 4} $ & $ 69.5 ^{+ 8.3} _{- 7.0} $  & $7 \times 10^{-4}$    &  \\ 
HD~15407 & 0.665   & 0.434 & $ 1.53 \pm 0.10 $   & $ 553 ^{+34} _{-31} $ & $  0.5 ^{+ 0.1} _{- 0.1} $  & $6 \times 10^{-3}$    &  \\ 
HD~39060 & --      & 4.288 & --                  & $106 ^{+5} _{-5} $ & $  20.4 ^{+ 2.1} _{- 1.8} $  & $2 \times 10^{-3}$    &    \\ 
HD~39415 & 0.159   & 0.256 & $ 0.62 \pm 0.08 $   & $ 321 ^{+17} _{-17} $ & $  1.6 ^{+ 0.2} _{- 0.2} $  & $6 \times 10^{-3}$    &  \\ 
HD~64145 & --      & 0.120 & --                  & ($<170$)              & ($>16.9$)                   & ($>9 \times 10^{-5}$) &     \\ 
HD~75809 & --      & 0.064 & --                  & ($<170$)              & ($>5.6$)                    & ($>2 \times 10^{-4}$) &  \\ 
HD~89125 & --      & 0.154 & --                  & ($<170$)              & ($>4.0$)                    & ($>3 \times 10^{-4}$) &  \\ 
HD~98800 & 0.726   & 6.076 & $ 0.12 \pm 0.01 $   & $ 190 ^{+ 3} _{- 3} $ & $  1.8 ^{+ 0.1} _{- 0.1} $  & $8 \times 10^{-2}$    &  \\ 
HD~105209 & 0.118   & 0.177 & $ 0.67 \pm 0.06 $   & $ 332 ^{+13} _{-13} $ & $  5.8 ^{+ 0.5} _{- 0.4} $  & $4 \times 10^{-4}$    &  \\ 
HD~106797 & --      & 0.097 & --                  & ($<170$)              & ($>17.5$)                   & ($>2 \times 10^{-4}$) &  \\ 
HD~109573 & 0.069   & 1.309 & $ 0.05 \pm 0.02 $   & $ 158 ^{+11} _{-13} $ & $ 15.3 ^{+ 2.9} _{- 1.9} $  & $2 \times 10^{-3}$    & Yes \\ 
HD~110058 & --      & 0.072 & --                  & $ 104 ^{+ 1} _{- 1} $ & $ 22.7 ^{+ 0.4} _{- 0.4} $  & $2 \times 10^{-3}$    &  \\ 
HD~113457 & --      & 0.080 & --                  & ($<170$)              & ($>22.2$)                   & ($>3 \times 10^{-4}$) &  \\ 
HD~113766 & 1.160   & 1.396 & $ 0.83 \pm 0.06 $   & $ 369 ^{+13} _{-13} $ & $  1.2 ^{+ 0.1} _{- 0.1} $  & $2 \times 10^{-2}$    &  \\ 
HD~121617 & 0.025   & 0.308 & $ 0.08 \pm 0.03 $   & $ 173 ^{+12} _{-14} $ & $ 21.5 ^{+ 3.9} _{- 2.7} $  & $2 \times 10^{-3}$    & Yes \\ 
HD~123356 & 0.258   & 0.319 & $ 0.81 \pm 0.17 $   & $ 365 ^{+38} _{-38} $ & $  0.7 ^{+ 0.2} _{- 0.1} $  & $8 \times 10^{-4}$    &  \\ 
HD~145263 & 0.194   & 0.477 & $ 0.41 \pm 0.04 $   & $ 272 ^{+ 8} _{- 9} $ & $  2.1 ^{+ 0.1} _{- 0.1} $  & $1 \times 10^{-2}$    &  \\ 
HD~165014 & 1.212   & 0.927 & $ 1.31 \pm 0.05 $   & $ 488 ^{+12} _{-13} $ & $  0.9 ^{+ 0.0} _{- 0.0} $  & $2 \times 10^{-2}$    &  \\ 
HD~166191 & 1.715   & 2.445 & $ 0.70 \pm 0.05 $   & $ 340 ^{+11} _{-12} $ & $  1.4 ^{+ 0.1} _{- 0.1} $  & $5 \times 10^{-2}$    & Yes \\ 
HD~167905 & 2.114   & 1.717 & $ 1.23 \pm 0.06 $   & $ 467 ^{+15} _{-15} $ & $  0.7 ^{+ 0.0} _{- 0.0} $  & $2 \times 10^{-2}$    &  \\ 
HD~175726 & --      & 0.125 & --                  & ($<170$)              & ($>2.9$)                    & ($>6 \times 10^{-4}$) &  \\ 
HD~176137 & 4.552   & 4.696 & $ 0.97 \pm 0.27 $   & $ 401 ^{+67} _{-61} $ & $  1.0 ^{+ 0.4} _{- 0.3} $  & $9 \times 10^{-3}$    &  \\ 
HD~181296 & --      & 0.206 & --                  & $ 123 ^{+ 5} _{- 5} $ & $ 24.1 ^{+ 2.1} _{- 1.8} $  & $2 \times 10^{-4}$    &  \\ 
\hline                                   
\end{tabular}
\end{table*}

\subsubsection{Lack of Cooler Material in Outer Region}

The flux contributed from the inner dust for each star is shown in Figure~\ref{SEDfit}. 
The excess component of our samples except for HD~109573, HD~121617, and HD~166191
can be fitted fairly well only with the inner dust component of a single blackbody. 
This implies the lack of cooler material in the outer region around those stars 
and truncation of the disk at inner radii, although FIR and sub-mm observations 
with high sensitivity are required to accurately investigate the presence of the outer material. 
Only one object (HD~166191) shows a rising SED towards $60~\micron$ among our 13 FGK-type samples, 
while about half of the samples of 11 A-type stars show FIR excess, 
suggesting that solar-type stars in our sample tend to have less cooler material in the outer region 
than A stars. 

\cite{rhee07} suggest that the dust temperature in typical debris disks discovered with {\it IRAS} observations
is less than 100~K and the peak of the excess flux density comes to around $60-100~\micron$. 
However, most of the debris disks detected by {\it AKARI}/IRC in this work have abundant warm dust, 
while cooler material in the outer regions is not conspicuous, 
suggesting that we detect a sub-group of the debris disks, in which warm dust grains are 
more abundant than in debris disks detected by {\it IRAS}.
Special mechanisms, such as dust trapping by the resonance perturbation of planets 
or generation of a large amount of warm dust, 
might play significant roles in the inner region of disk in these systems.

\subsubsection{Fractional Luminosity of Inner Disk}

Estimating the amount of dust is important to characterise the debris disk. 
One of the observable indicators of the dust abundance is its fractional luminosity, 
$f=L_{\rm dust}/L_{*}$, i.e., the IR luminosity from the disk divided by the stellar luminosity.
We derive the fractional luminosities ($f_{\rm obs}$) of the inner debris disk 
by integrating the intrinsic stellar emission and the observed excess emission of the inner debris component. 
The derived fractional luminosities are shown in Table~\ref{inner}.
It should be noted that the fractional luminosities for the inner disks with the cold dust component
should be regarded as upper limits, 
since we do not take account of the contribution from the cold dust. 
The derived fractional luminosities range from $\sim 10^{-4}$ to $\sim 10^{-1}$. 
The fractional luminosity of our own zodiacal cloud is estimated to be of the order $\sim 10^{-7}$ \citep{backman93}. 
Thus the debris disks in our sample are typically by more than 1000 times brighter than our own zodiacal cloud. 

\cite{wyatt07} developed a simple model for the steady-state evolution of debris disks 
produced by collisions and suggested that the maximum fractional luminosity ($f_{\rm max}$) 
can be written by 
\begin{eqnarray}
f_{\rm max} = 0.16 \times 10^{-3} \ R_{\rm dust}^{7/3} \ M_*^{-5/6} \ L_*^{-0.5} \ t_{\rm age}^{-1}
\end{eqnarray}
for the fixed model parameters
(belt width: $dr/r = 0.5$; planetesimal strength: $Q^*_D = 200$~J/kg; 
planetesimal eccentricity: $e = 0.05$; and diameter of the largest planetesimal in cascade: 
$D_{\rm c} = 2000$~km).
The values of $f_{\rm max}$ were calculated using the parameters of the central star listed in Table~\ref{stellar} 
and the estimated parameters for the inner disk component in Table~\ref{inner}.  

The ratios of the observed fractional luminosities to 
the theoretical maximum fractional luminosities ($f_{\rm obs}/f_{\rm max}$) 
are summarised in Table~\ref{fobsfmax}. 
Some debris disks show fractional luminosities of $f_{\rm obs}/f_{\rm max} \gg 100$, 
suggesting that these debris disks cannot be accounted for by simple models of the steady-state debris disks, 
and that transient events likely play a significant role, even taking account of the uncertainties in the models \citep{wyatt07}. 

One of the transient phenomena producing a large amount of debris dust 
around a star is a dynamical instability that scatters planetesimals inward 
from a more distant planetesimal belt. 
Dust is released from unstable planetesimals following collisions and sublimation. 
This is akin to the late heavy bombardment (LHB) in the solar system, 
the cataclysmic event that occurred about 700~Myr after the initial formation of the solar system, 
as implied by Moon's cratering record \citep[e.g.][]{hartmann00}.

To develop a LHB-like event around a star, 
a newly born planet would have to migrate and stir up planetesimals. 
Since planet formation is expected to last for several hundred Myr around solar-type stars, 
LHB-like events could be the cause of dense warm debris disks 
around the young (age $\lesssim$ Gyr) FGK systems of HD~113766, HD~145263, and HD~98800. 
On the other hand, HD~89125, HD~15407, and HD~175726 are older than one Gyr 
and planet formation is expected to have already been finished. 
Therefore another dust production mechanism would be required
to account for the large fractional luminosities of HD~89125, HD~15407, and HD~175726. 
It is worth noting that recent FIR observations of HD~15407 by {\it Herschel} and {\it AKARI} revealed 
the absence of cold dust around the star, not supporting an LHB-like event as the origin of 
the bright debris disk around the star \citep{fujiwara12b}.


\begin{table}
\caption{Comparison of observed ($f_{\rm obs}$) and theoretical fractional luminosities ($f_{\rm max}$) 
for our debris disk sample whose age is available. \label{fobsfmax}}
\centering                          
\begin{tabular}{ccccc}        
\hline\hline                 
Name      & $f_{\rm obs}$ & $f_{\rm max}$ & $f_{\rm obs}/f_{\rm max}$ \\
\hline                        
HD~3003   & $7 \times 10^{-5}$ & $7 \times 10^{-5}$ & 1 \\
HD~9672   & $7 \times 10^{-4}$ & $2 \times 10^{-3}$ & 0.4 \\
HD~15407  & $6 \times 10^{-3}$ & (300--1)~$\times 10^{-8}$ & 20000--600000 \\
HD~39060  & $2 \times 10^{-3}$ & $3 \times 10^{-3}$ & 0.9 \\
HD~89125  & ($>3 \times 10^{-4}$) & ($>7 \times 10^{-7}$) & ($>500$) \\
HD~98800  & $8 \times 10^{-2}$ & $6 \times 10^{-5}$ & 1000 \\
HD~106797 & ($>2 \times 10^{-4}$) & ($>1 \times 10^{-3}$) & ($>0.2$) \\
HD~109573 & $2 \times 10^{-3}$ & $5 \times 10^{-5}$ & 40 \\
HD~110058 & $2 \times 10^{-3}$ & $2 \times 10^{-3}$ & 0.7 \\
HD~113457 & ($>3 \times 10^{-4}$) & ($>2 \times 10^{-3}$) & ($>0.2$) \\
HD~113766 & $2 \times 10^{-2}$ & $7 \times 10^{-6}$ & 3000 \\
HD~145263 & $1 \times 10^{-2}$ & $4 \times 10^{-5}$ & 300 \\
HD~175726 & ($>6 \times 10^{-4}$) & ($>$ (40--1)~$\times 10^{-7}$) & ($>$ 200--6000) \\
HD~181296 & $2 \times 10^{-4}$ & $1 \times 10^{-3}$ & 0.2 \\
\hline                                   
\end{tabular}
\end{table}

\subsubsection{Difference between A and FGK stars}

\begin{figure*}
\centering
\includegraphics[width=4.4cm]{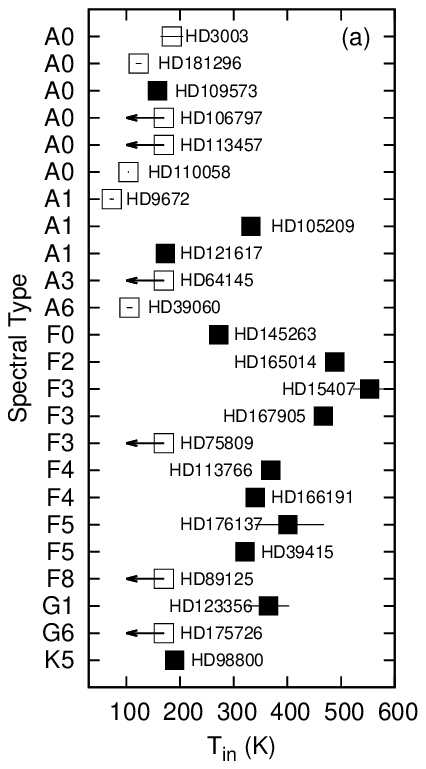}
\includegraphics[width=4.4cm]{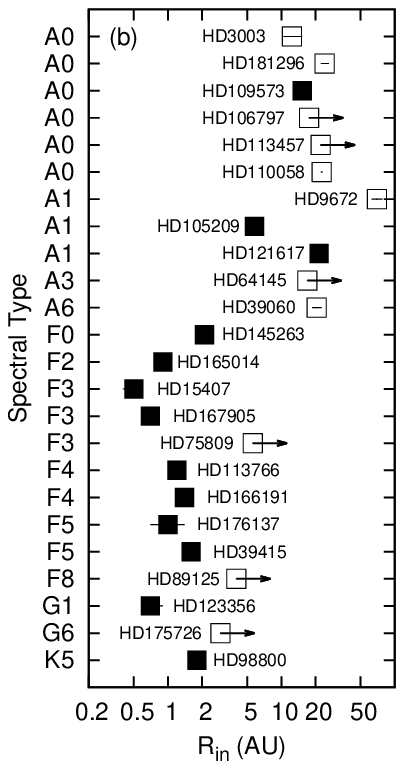}
\includegraphics[width=4.4cm]{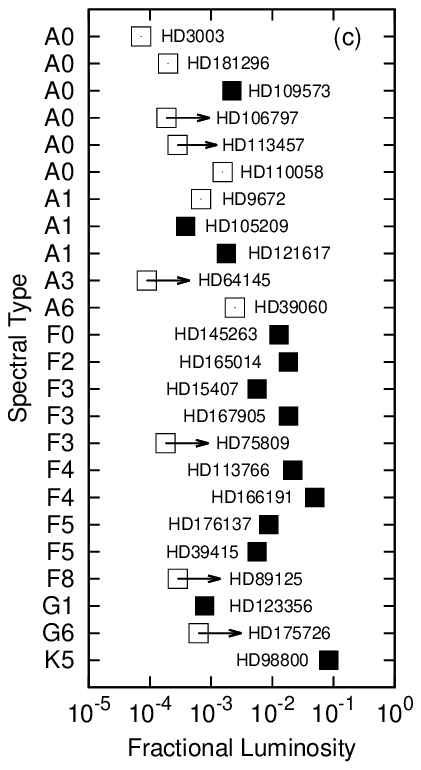}
\includegraphics[width=4.4cm]{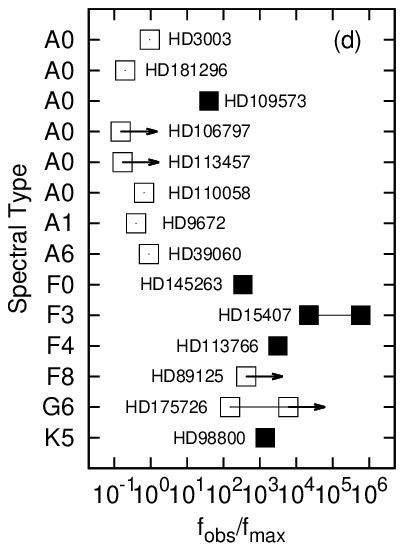}
\caption{Distributions of $T_{\rm in}$ (a), $R_{\rm in}$ (b), 
fractional luminosity (c) and $f_{\rm obs}/f_{\rm max}$ (d). 
Filled and open squares indicate the stars with and without $9~\micron$ excess, respectively. 
We set $T_{\rm in}$ at 170~K for the stars without $9~\micron$ excess and photometric information at $\lambda \gtrsim 24~\micron$.
\label{vsSpT}}
\end{figure*}

To investigate the relationship between the debris disks and the spectral type of the central stars,
we plot the distributions of $T_{\rm in}$, $R_{\rm in}$, fractional luminosity 
and $f_{\rm obs}/f_{\rm max}$ sorting by the spectral types in Figure~\ref{vsSpT}. 
It is seen that all of the debris disks investigated here change their characteristics appreciably 
around the spectral type F0; 
$T_{\rm in} \lesssim 200$~K for A stars while $T_{\rm in} \gtrsim 300$~K for FGK stars; 
$R_{\rm in} \gtrsim $ a few AU for A stars while $R_{\rm in} \lesssim $ a few AU for FGK stars; 
the fractional luminosities for A stars are distributed around $10^{-5}-10^{-3}$ 
while those for FGK stars are in a range $10^{-3}-10^{-1}$, 
and $f_{\rm obs}/f_{\rm max}$ for A stars is distributed around unity 
while that for FGK stars is $\gtrsim 100$. 
In other words, the properties of the inner disks of A and FGK stars are different from each other. 
Most of our debris disk samples around FGK stars possess warm dust 
without an appreciable amount of cooler dust, 
while those debris disks with $18~\micron$ excesses around A stars show little excess at $9~\micron$. 
Our debris disk samples around FGK stars may belong to a sub-group of the debris disks 
in which violent dust supply is taking place in the inner region ($\lesssim$ a few~AU) of the disk. 

Radiation pressure is capable of removing dust grains 
and changing the radial distribution of dust. 
When the radiation-pressure-to-gravity ratio $\beta=F_{\rm rad}/F_{\rm grav}$ 
becomes larger than unity, the outward force dominates and the dust grains are blown out. 
The ratio $\beta$ can be approximated by
\begin{eqnarray}
\beta=(0.4~\micron/D)(2.7~{\rm g}/{\rm cm}^{3}/\rho)(L_*/M_*), 
\end{eqnarray}
where $D$ is the grain size, $\rho$ is the grain density, 
and $L_*$ and $M_*$ are in units of $L_\odot$ and $M_\odot$ \citep{burns79}. 
For $\micron$-sized ($D \sim 1~\micron$) silicate ($\rho \sim 2.7~{\rm g}/{\rm cm}^{3}$) grains, 
$\beta$ becomes about unity around a F0 star ($M_*=1.6M_\odot$ and $L_*=4.1L_\odot$). 
Therefore $\micron$-sized grains may be blown out by radiation pressure 
in the vicinity of A stars, while they can survive in FGK stars. 

\cite{morales09} observed 52 main-sequence A and late B type stars with the {\it Spitzer}/IRS
that showed excess emission at $24~\micron$ in the {\it Spitzer}/MIPS observations. 
They found no prominent spectral features evident in any of the spectra, 
suggesting that fine particles that show spectral features 
are absent around early-type stars. 

By taking the estimated $\beta$ value and the absence of small dust grains around early-type stars into account, 
it is suggested that $\micron$-sized grains are blown out by radiation pressure from A stars
and that the difference of the disk characteristics between A and FGK stars seen in our sample may be driven by the radiation pressure on grains.

\subsection{Individual Star}

Here we present notes about the individual debris disk candidates with $18~\micron$ excess discovered with {\it AKARI}  
or confirmed in this work for the first time after the report by \cite{oudmaijer92}.

\subsubsection{HD~15407}

HD~15407 is an F3V star at $d=55$~pc, 
whose IR excess was originally suggested by \cite{oudmaijer92}  
and confirmed with our present {\it AKARI} study. 
The SIMBAD database indicates HD~15407 is a star in double system with the nearby K2V star HIP~11692 (HD~15407B) 
with the apparent separation of $21\farcs2$. 
In this paper we refer the primary star (frequently called as HD~15407A) simply as HD~15407.
HD~15407 star may be mature and the age is estimated as 2.1~Gyr by \cite{holmberg09}. 
On the other hand, a recent paper \citep{melis10} suggests 
a much younger age ($\sim 80$~Myr) based on the lithium 6710 \AA \ absorption and the motion in the Galaxy. 
This star is one of the most peculiar debris disk candidates in our sample 
since the fractional luminosity is much larger than the steady-state evolution model of planetesimals 
($f_{\rm obs}/f_{\rm max}> 10^5$). 
In addition, abundant silica (SiO$_2$) dust, which is rare dust species in the universe, 
is seen in the {\it Spitzer}/IRS spectrum \citep{fujiwara12a,olofsson12}. 
Special mechanisms to generate a large amount of peculiar dust 
play a significant role around the star.

\subsubsection{HD~39415}

HD~39415 is an F5V star, 
whose IR excess was originally suggested by \cite{oudmaijer92} 
and confirmed with our present {\it AKARI} study. 
No age information is available at present. 
\cite{holmes03} report non-detection of sub-mm emission at 870 and $1300~\micron$ 
($3\sigma$ limits are 18.75 and 10.68~mJy at 870 and $1300~\micron$, respectively)
with the Heinrich Hertz Telescope at the Submillimeter Telescope Observatory, 
suggesting that the amount of cool material around the star, if any, is quite small.

\subsubsection{HD~64145}

HD~64145 is an A3V star at $d=78$~pc, 
whose $18~\micron$ excess is found in this {\it AKARI} sample. 
No age information is available at present. 
This star is reported as a binary system with a secondary on a 582-day orbit \cite[e.g.][]{abt05}. 
Although no excess at $9~\micron$ is detected with the {\it AKARI}/IRC measurements, 
the {\it IRAS} $12~\micron$ flux density is slightly higher than the photospheric level. 
Since the inner radius of the disk is derived as $R_{\rm in} \sim 17$~AU and thus 
an apparent separation of $0\farcs22$ is expected, 
an extended disk structure might be resolved spatially in the scattered light 
with a high contrast coronagraph with adaptive optics on 8--10-m class ground-base telescopes. 
It should be noted that this star is contained in the {\it WISE} 
All-Sky Source Catalogue \citep{cutri12} as J075329.80+264556.5. 
The flux density at $22~\micron$ (W4 band) is estimated as $\sim 130$~mJy 
in the {\it WISE} All-Sky Source Catalogue and is consistent with the expected photoshperic level, 
suggesting no significant excess at $22~\micron$. 
Photometries with narrow/intermediate filters at around $20~\micron$ are required for examination of the inconsistency 
in the flux density between {\it AKARI} and {\it WISE}.

\subsubsection{HD~75809}

HD~75809 is an F3V star, 
whose $18~\micron$ excess is found in this {\it AKARI} sample. 
No age information is available at present. 
No $9~\micron$ excess is found with the {\it AKARI} observations, 
suggesting that the temperature of the debris disk is at the lowest end amongst our samples.

\subsubsection{HD~89125}

HD~89125 is a nearby F8V star at $d=23$~pc, 
whose $18~\micron$ excess is found in this {\it AKARI} sample. 
This star is mature and the age is estimated as 6.5~Gyr \citep{takeda07}. 
The fractional luminosity of the star is much larger than the steady-state evolution model of planetesimals 
($f_{\rm obs}/f_{\rm max}> 500$), suggesting that transient dust production events might 
play a significant role around the star. 
It should be noted that the star is reported as being metal poor ([Fe/H]$=-0.39$) 
and one of the most metal-poor debris systems among the known sample \citep{beichman06}. 
This star is contained in the {\it WISE} 
All-Sky Source Catalogue \citep{cutri12} as J101714.20+230621.6. 
The flux density at $22~\micron$ (W4 band) is estimated as $\sim 140$~mJy 
in the {\it WISE} All-Sky Source Catalogue and is consistent with the expected photoshperic level, 
suggesting no significant excess at $22~\micron$. 
Photometry with narrow/intermediate filters at around $20~\micron$ is required for examination of the inconsistency.

\subsubsection{HD~106797}

HD~106797 is an A0V star at $d=99$~pc, 
whose $18~\micron$ excess is found in this {\it AKARI} sample. 
\cite{dezeeuw99} classified HD~106797 as a member of 
Lower Centaurus Crux group, whose age is estimated as $10-20$~Myr. 
\cite{fujiwara09} suggested the presence of narrow spectral features 
between 11 and $12~\micron$, which is attributable to (sub-)$\micron$-sized crystalline silicates 
based on the multi-band photometric data collected in the follow-up observations with Gemini/T-ReCS. 

\subsubsection{HD~113457}

HD~113457 is an A0V star at $d=95$~pc, 
whose $18~\micron$ excess is found in this {\it AKARI} sample. 
\cite{dezeeuw99} classifies HD~113457 as a member of 
Lower Centaurus Crux group, whose age is estimated as $10-20$~Myr. 
Since the inner radius of the disk is estimated to be $R_{\rm in} \sim 22$~AU and thus 
an apparent separation of $0\farcs23$ is expected, 
an extended disk structure might be resolved spatially in the scattered light 
with a high contrast coronagraph with adaptive optics.

\subsubsection{HD~165014}

HD~165014 is an F2V star, 
whose 9 and $18~\micron$ excesses are found in this {\it AKARI} sample. 
No age information is available at present. 
This star was also detected at 8--14$~\micron$ with {\it MSX} 
and possible excesses at the wavelengths have been suggested by \cite{clarke05}.
Strong dust features attributable to crystalline enstatite (MgSiO$_3$) 
are seen in its {\it Spitzer}/IRS spectrum \citep{fujiwara10}
while crystalline forsterite (Mg$_2$SiO$_4$) features, which is typically more abundant around young stars, 
are not seen. 
Possible formation of enstatite dust from differentiated parent bodies is suggested according to the solar system analog. 
Special mechanisms to generate a large amount of crystalline enstatite dust
must play a role around the star \citep{fujiwara10}.

\subsubsection{HD~175726}

HD~175726 is a nearby G6V star at $d=27$~pc, 
whose $18~\micron$ excess is found in this {\it AKARI} sample. 
The stellar age is estimated as $4.8 \pm 3.5$~Gyr by \cite{bruntt09}. 
The fractional luminosity is larger than the steady-state evolution model of planetesimals 
($f_{\rm obs}/f_{\rm max}> 10^3$), suggesting that some transient dust production events  
play a role around the star. 
Since the stellar age is in a range during which LHB took place in the solar system, 
the bright debris disk around HD~175726 might be produced by a LHB-like phenomenon. 
It should be noted that this star is contained in the {\it WISE} 
All-Sky Source Catalogue \citep{cutri12} as J185637.17+041553.6. 
Although the flux density at $22~\micron$ (W4 band) is estimated as $\sim 110$~mJy 
in the {\it WISE} Catalogue and suggests excess at the wavelength, 
it is considerably lower than the {\it AKARI} measurement at $18~\micron$. 
The discrepancy may suggest temporal variability of MIR excess towards HD~175726 over a few years. 
A recent study reveals a dramatic decrease in the infrared excess seen in one of the warm dust debris candidates, 
the young Sun-like star TYC 8241-2652-1, in several years, 
which suggests removal of warm debris from the system in a very short time scale \citep{melis12}. 
Monitoring observations of HD~175726 around $20~\micron$ could reveal the nature and evolution of its debris system. 
The star is a CoRoT asteroseismic target and 
the presence of weak solar-like oscillations is reported \citep{mosser09}
while no hint of planets orbiting the star is so far suggested.

\subsubsection{HD~176137}

HD~176137 is an F5 star, 
whose $18~\micron$ excess is found in this {\it AKARI} sample. 
No age information is available at present. 
This star is located in the direction towards a highly obscured region in the Galaxy, 
whose interstellar extinction is estimated as $A_V =  27.1$ 
by ``Galactic Dust Reddening and Extinction'' at the IRSA\footnote{URL: http://irsa.ipac.caltech.edu/applications/DUST/.}. 
The extinction towards the star derived from our fitting procedure of photosphere 
is $A_v =  13.0$, much smaller than the estimated extinction, 
suggesting that the star is located near to us. 
HD~176137 is a double star system with a separation of $0\farcs5$--$1\arcsec$ (at the epoch of 1910-2009) 
as resolved by speckle interferometry \citep{douglass00}. 
The nature of the nearby star is not known enough. 
Ground-based MIR observations of the star with high spatial resolution are needed for further discussion on the possible confusion 
in the excess from the nearby star.

%

%

\section{Summary}

We report the results of a systematic survey of warm ($T \gtrsim 150$~K) debris disks 
based on photometric measurements at $18~\micron$ take from the {\it AKARI}/IRC All-Sky Survey data. 
We have found 24 debris disk candidates with bright MIR excess emission 
above the stellar photospheric emission out of 856 sources that were detected at $18~\micron$. 
Among them 8 stars are newly discovered in this work. 
We found that 13 stars of the 24 debris disk candidates also show excess emission at $9~\micron$. 
The overall apparent debris disk frequency is derived as $2.8 \pm 0.6$\% for our sample. 
We have identified a tendency for the frequency to increase towards earlier type stars. 

The temperature, radius, and fractional luminosity of the inner disk component of the candidates 
are derived from the 9 to $18~\micron$ flux densities of the excess emission sources. 
We find that A stars and solar-type FGK stars have different characteristics of their debris disks; 
most of those around FGK stars possess warm dust without any cool dust component, 
while the debris disks around A stars show little evidence for excess at $9~\micron$ 
and appear to have lower temperature ($T \lesssim 200$~K) debris dust. 
In addition, considering only the objects for which the age information
is available, we find that the fractional luminosities of the inner
debris disks around FGK stars cannot be explained by steady-state
evolutionary models of debris disks that are sustained by 
collisions of planetesimals. On the other hand, the debris disks we have
detected around A stars are found to fit well with the steady-state model. 
We propose that the debris disks around FGK stars belong to a sub-group of debris disks,  
in which violent dust supply and processing are taking place in the inner region ($\lesssim$ a few~AU) of the disk. 
The suggested difference in the debris disk characteristics between A and solar-type FGK stars 
can be explored by the next version of the {\it AKARI}/IRC PSC 
and the data from {\it WISE}, 
which will dramatically improve the sensitivity of MIR photometry and should allow us 
to study a larger population of these very interesting objects.

\begin{acknowledgements}
This research is based on observations with the {\it AKARI}, 
a JAXA project with the participation of ESA 
and on data collected at the Subaru Telescope and the Gemini Observatory. 
It has made use of the SIMBAD database and the VizieR catalogue access tool, CDS, Strasbourg, France 
and the data products from 2MASS. H.F. and S.T. was financially supported by the Japan Society for the Promotion of Science. 
This research was supported by KAKENHI (07J02823 and 23103002).
\end{acknowledgements}


\bibliographystyle{aa} 
\bibliography{references.bib} 

\begin{thebibliography}{84}
\expandafter\ifx\csname natexlab\endcsname\relax\def\natexlab#1{#1}\fi

\bibitem[{{Abia} {et~al.}(2009){Abia}, {de Laverny}, {Recio-Blanco},
  {Dom{\'{\i}}nguez}, {Cristallo}, \& {Straniero}}]{abia09}
{Abia}, C., {de Laverny}, P., {Recio-Blanco}, A., {et~al.} 2009, Publications
  of the Astronomical Society of Australia, 26, 351

\bibitem[{{Abt}(2005)}]{abt05}
{Abt}, H.~A. 2005, \apj, 629, 507

\bibitem[{{Acke} {et~al.}(2009){Acke}, {Min}, {van den Ancker}, {Bouwman},
  {Ochsendorf}, {Juhasz}, \& {Waters}}]{acke09}
{Acke}, B., {Min}, M., {van den Ancker}, M.~E., {et~al.} 2009, \aap, 502, L17

\bibitem[{{Aumann} {et~al.}(1984){Aumann}, {Beichman}, {Gillett}, {de Jong},
  {Houck}, {Low}, {Neugebauer}, {Walker}, \& {Wesselius}}]{aumann84}
{Aumann}, H.~H., {Beichman}, C.~A., {Gillett}, F.~C., {et~al.} 1984, \apjl,
  278, L23

\bibitem[{{Backman} \& {Paresce}(1993)}]{backman93}
{Backman}, D.~E. \& {Paresce}, F. 1993, in Protostars and Planets III, ed.
  {E.~H.~Levy \& J.~I.~Lunine}, 1253--1304

\bibitem[{{Bastian} \& {R{\"o}ser}(1993)}]{bastian93}
{Bastian}, U. \& {R{\"o}ser}, S. 1993, {Catalog of Positions and Proper
  Motions: South (Heidelberg: Astron. Rechen Inst.)}, ed. S.~Bastian, U.
  \&~R{\"o}ser

\bibitem[{{Beichman} {et~al.}(2006){Beichman}, {Bryden}, {Stapelfeldt},
  {Gautier}, {Grogan}, {Shao}, {Velusamy}, {Lawler}, {Blaylock}, {Rieke},
  {Lunine}, {Fischer}, {Marcy}, {Greaves}, {Wyatt}, {Holland}, \&
  {Dent}}]{beichman06}
{Beichman}, C.~A., {Bryden}, G., {Stapelfeldt}, K.~R., {et~al.} 2006, \apj,
  652, 1674

\bibitem[{{Beichman} {et~al.}(1988){Beichman}, {Neugebauer}, {Habing}, {Clegg},
  \& {Chester}}]{beichman88}
{Beichman}, C.~A., {Neugebauer}, G., {Habing}, H.~J., {Clegg}, P.~E., \&
  {Chester}, T.~J., eds. 1988, {Infrared astronomical satellite (IRAS) catalogs
  and atlases. Volume 1: Explanatory supplement}, Vol.~1

\bibitem[{{Bruntt}(2009)}]{bruntt09}
{Bruntt}, H. 2009, \aap, 506, 235

\bibitem[{{Burns} {et~al.}(1979){Burns}, {Lamy}, \& {Soter}}]{burns79}
{Burns}, J.~A., {Lamy}, P.~L., \& {Soter}, S. 1979, Icarus, 40, 1

\bibitem[{{Chen} {et~al.}(2005){Chen}, {Patten}, {Werner}, {Dowell},
  {Stapelfeldt}, {Song}, {Stauffer}, {Blaylock}, {Gordon}, \&
  {Krause}}]{chen05}
{Chen}, C.~H., {Patten}, B.~M., {Werner}, M.~W., {et~al.} 2005, \apj, 634, 1372

\bibitem[{{Clarke} {et~al.}(2005){Clarke}, {Oudmaijer}, \&
  {Lumsden}}]{clarke05}
{Clarke}, A.~J., {Oudmaijer}, R.~D., \& {Lumsden}, S.~L. 2005, \mnras, 363,
  1111

\bibitem[{{Cohen} {et~al.}(1999){Cohen}, {Walker}, {Carter}, {Hammersley},
  {Kidger}, \& {Noguchi}}]{cohen99}
{Cohen}, M., {Walker}, R.~G., {Carter}, B., {et~al.} 1999, \aj, 117, 1864

\bibitem[{{Cox}(2000)}]{cox00}
{Cox}, A.~N. 2000, {Allen's astrophysical quantities}, ed. A.~N. Cox

\bibitem[{{Crifo} {et~al.}(1997){Crifo}, {Vidal-Madjar}, {Lallement}, {Ferlet},
  \& {Gerbaldi}}]{crifo97}
{Crifo}, F., {Vidal-Madjar}, A., {Lallement}, R., {Ferlet}, R., \& {Gerbaldi},
  M. 1997, \aap, 320, L29

\bibitem[{{Cutri} {et~al.}(2003){Cutri}, {Skrutskie}, {van Dyk}, {Beichman},
  {Carpenter}, {Chester}, {Cambresy}, {Evans}, {Fowler}, {Gizis}, {Howard},
  {Huchra}, {Jarrett}, {Kopan}, {Kirkpatrick}, {Light}, {Marsh}, {McCallon},
  {Schneider}, {Stiening}, {Sykes}, {Weinberg}, {Wheaton}, {Wheelock}, \&
  {Zacarias}}]{cutri03}
{Cutri}, R.~M., {Skrutskie}, M.~F., {van Dyk}, S., {et~al.} 2003, {2MASS All
  Sky Catalog of point sources.}, ed. {Cutri, R.~M., Skrutskie, M.~F., van Dyk,
  S., Beichman, C.~A., Carpenter, J.~M., Chester, T., Cambresy, L., Evans, T.,
  Fowler, J., Gizis, J., Howard, E., Huchra, J., Jarrett, T., Kopan, E.~L.,
  Kirkpatrick, J.~D., Light, R.~M., Marsh, K.~A., McCallon, H., Schneider, S.,
  Stiening, R., Sykes, M., Weinberg, M., Wheaton, W.~A., Wheelock, S., \&
  Zacarias, N.}

\bibitem[{{Cutri} {et~al.}(2012){Cutri}, {Wright}, {Conrow}, {Bauer},
  {Benford}, {Brandenburg}, {Dailey}, {Eisenhardt}, {Evans}, {Fajardo-Acosta},
  {Fowler}, {Gelino}, {Grillmair}, {Harbut}, {Hoffman}, {Jarrett},
  {Kirkpatrick}, {Leisawitz}, {Liu}, {Mainzer}, {Marsh}, {Masci}, {McCallon},
  {Padgett}, {Ressler}, {Royer}, {Skrutskie}, {Stanford}, {Wyatt}, {Tholen},
  {Tsai}, {Wachter}, {Wheelock}, {Yan}, {Alles}, {Beck}, {Grav}, {Masiero},
  {McCollum}, {McGehee}, {Papin}, \& {Wittman}}]{cutri12}
{Cutri}, R.~M., {Wright}, E.~L., {Conrow}, T., {et~al.} 2012, {Explanatory
  Supplement to the WISE All-Sky Data Release Products}, Tech. rep.

\bibitem[{{de Zeeuw} {et~al.}(1999){de Zeeuw}, {Hoogerwerf}, {de Bruijne},
  {Brown}, \& {Blaauw}}]{dezeeuw99}
{de Zeeuw}, P.~T., {Hoogerwerf}, R., {de Bruijne}, J.~H.~J., {Brown}, A.~G.~A.,
  \& {Blaauw}, A. 1999, \aj, 117, 354

\bibitem[{{Douglass} {et~al.}(2000){Douglass}, {Mason}, {Rafferty},
  {Holdenried}, \& {Germain}}]{douglass00}
{Douglass}, G.~G., {Mason}, B.~D., {Rafferty}, T.~J., {Holdenried}, E.~R., \&
  {Germain}, M.~E. 2000, \aj, 119, 3071

\bibitem[{{Egan} {et~al.}(2003){Egan}, {Price}, {Kraemer}, {Mizuno}, {Carey},
  {Wright}, {Engelke}, {Cohen}, \& {Gugliotti}}]{eagan03}
{Egan}, M.~P., {Price}, S.~D., {Kraemer}, K.~E., {et~al.} 2003, VizieR Online
  Data Catalog, 5114, 0

\bibitem[{{Eiroa} {et~al.}(2011){Eiroa}, {Marshall}, {Mora}, {Krivov},
  {Montesinos}, {Absil}, {Ardila}, {Ar{\'e}valo}, {Augereau}, {Bayo}, {Danchi},
  {Del Burgo}, {Ertel}, {Fridlund}, {Gonz{\'a}lez-Garc{\'{\i}}a}, {Heras},
  {Lebreton}, {Liseau}, {Maldonado}, {Meeus}, {Montes}, {Pilbratt}, {Roberge},
  {Sanz-Forcada}, {Stapelfeldt}, {Th{\'e}bault}, {White}, \& {Wolf}}]{eiroa11}
{Eiroa}, C., {Marshall}, J.~P., {Mora}, A., {et~al.} 2011, \aap, 536, L4

\bibitem[{{Fitzpatrick} \& {Massa}(2009)}]{fitzpatrick09}
{Fitzpatrick}, E.~L. \& {Massa}, D. 2009, \apj, 699, 1209

\bibitem[{{Fricke} {et~al.}(1991){Fricke}, {Schwan}, {Corbin}, {Bastian},
  {Bien}, {Cole}, {Jackson}, {J{\"a}hrling}, {Jahrei{\ss}}, {Lederle}, \&
  {R{\"o}ser}}]{fricke91}
{Fricke}, W., {Schwan}, H., {Corbin}, T., {et~al.} 1991, Veroeffentlichungen
  des Astronomischen Rechen-Instituts Heidelberg, 33, 1

\bibitem[{{Fricke} {et~al.}(1988){Fricke}, {Schwan}, {Lederle}, {Bastian},
  {Bien}, {Burkhardt}, {Du Mont}, {Hering}, {J{\"a}hrling}, {Jahrei{\ss}},
  {R{\"o}ser}, {Schwerdtfeger}, \& {Walter}}]{fricke88}
{Fricke}, W., {Schwan}, H., {Lederle}, T., {et~al.} 1988, Veroeffentlichungen
  des Astronomischen Rechen-Instituts Heidelberg, 32, 1

\bibitem[{{Fujiwara} {et~al.}(2010){Fujiwara}, {Onaka}, {Ishihara},
  {Yamashita}, {Fukagawa}, {Nakagawa}, {Kataza}, {Ootsubo}, \&
  {Murakami}}]{fujiwara10}
{Fujiwara}, H., {Onaka}, T., {Ishihara}, D., {et~al.} 2010, \apjl, 714, L152

\bibitem[{{Fujiwara} {et~al.}(2012{\natexlab{a}}){Fujiwara}, {Onaka}, {Takita},
  {Yamashita}, {Fukagawa}, {Ishihara}, {Kataza}, \& {Murakami}}]{fujiwara12b}
{Fujiwara}, H., {Onaka}, T., {Takita}, S., {et~al.} 2012{\natexlab{a}}, \apjl,
  759, L18

\bibitem[{{Fujiwara} {et~al.}(2012{\natexlab{b}}){Fujiwara}, {Onaka},
  {Yamashita}, {Ishihara}, {Kataza}, {Fukagawa}, {Takeda}, \&
  {Murakami}}]{fujiwara12a}
{Fujiwara}, H., {Onaka}, T., {Yamashita}, T., {et~al.} 2012{\natexlab{b}},
  \apjl, 749, L29

\bibitem[{{Fujiwara} {et~al.}(2009){Fujiwara}, {Yamashita}, {Ishihara},
  {Onaka}, {Kataza}, {Ootsubo}, {Fukagawa}, {Marshall}, {Murakami}, {Nakagawa},
  {Hirao}, {Enya}, \& {White}}]{fujiwara09}
{Fujiwara}, H., {Yamashita}, T., {Ishihara}, D., {et~al.} 2009, \apjl, 695, L88

\bibitem[{{Gielen} {et~al.}(2008){Gielen}, {van Winckel}, {Min}, {Waters}, \&
  {Lloyd Evans}}]{gielen08}
{Gielen}, C., {van Winckel}, H., {Min}, M., {Waters}, L.~B.~F.~M., \& {Lloyd
  Evans}, T. 2008, \aap, 490, 725

\bibitem[{{Hartmann} {et~al.}(2000){Hartmann}, {Ryder}, {Dones}, \&
  {Grinspoon}}]{hartmann00}
{Hartmann}, W.~K., {Ryder}, G., {Dones}, L., \& {Grinspoon}, D. 2000, {The
  Time-Dependent Intense Bombardment of the Primordial Earth/Moon System}, ed.
  R.~K. . e.~a. Canup, R.~M., 493--512

\bibitem[{{Hern{\'a}ndez} {et~al.}(2005){Hern{\'a}ndez}, {Calvet}, {Hartmann},
  {Brice{\~n}o}, {Sicilia-Aguilar}, \& {Berlind}}]{hernandeez05}
{Hern{\'a}ndez}, J., {Calvet}, N., {Hartmann}, L., {et~al.} 2005, \aj, 129, 856

\bibitem[{{Holmberg} {et~al.}(2009){Holmberg}, {Nordstr{\"o}m}, \&
  {Andersen}}]{holmberg09}
{Holmberg}, J., {Nordstr{\"o}m}, B., \& {Andersen}, J. 2009, \aap, 501, 941

\bibitem[{{Holmes} {et~al.}(2003){Holmes}, {Butner}, {Fajardo-Acosta}, \&
  {Rebull}}]{holmes03}
{Holmes}, E.~K., {Butner}, H.~M., {Fajardo-Acosta}, S.~B., \& {Rebull}, L.~M.
  2003, \aj, 125, 3334

\bibitem[{{Honda} {et~al.}(2004){Honda}, {Kataza}, {Okamoto}, {Miyata},
  {Yamashita}, {Sako}, {Fujiyoshi}, {Ito}, {Okada}, {Sakon}, \&
  {Onaka}}]{honda04}
{Honda}, M., {Kataza}, H., {Okamoto}, Y.~K., {et~al.} 2004, \apjl, 610, L49

\bibitem[{{Houk}(1978)}]{houk78}
{Houk}, N. 1978, {Michigan Catalog of Two-dimensional Spectral Types for HD
  Stars, Vol.\ 2 (Ann Arbor: Univ.\ Michigan Dept.\ Astron.)}, ed. N.~Houk

\bibitem[{{Houk}(1982)}]{houk82}
{Houk}, N. 1982, {Michigan Catalog of Two-dimensional Spectral Types for HD
  Stars, Vol.\ 3 (Ann Arbor: Univ.\ Michigan Dept.\ Astron.)}, ed. N.~Houk

\bibitem[{{Houk} \& {Cowley}(1975)}]{houk75}
{Houk}, N. \& {Cowley}, A.~P. 1975, {Michigan Catalog of Two-dimensional
  Spectral Types for HD Stars, Vol.\ 1 (Ann Arbor: Univ.\ Michigan Dept.\
  Astron.)}, ed. A.~P. Houk, N. \&~Cowley

\bibitem[{{Houk} \& {Smith-Moore}(1988)}]{houk88}
{Houk}, N. \& {Smith-Moore}, M. 1988, {Michigan Catalog of Two-dimensional
  Spectral Types for HD Stars, Vol.\ 4 (Ann Arbor: Univ.\ Michigan Dept.\
  Astron.)}, ed. J.~Warren, W.~H.

\bibitem[{{Houk} \& {Swift}(1999)}]{houk99}
{Houk}, N. \& {Swift}, C. 1999, {Michigan Catalog of Two-dimensional Spectral
  Types for HD Stars, Vol.\ 5 (Ann Arbor: Univ.\ Michigan Dept.\ Astron.)}, ed.
  C.~Houk, N. \&~Swift

\bibitem[{{Ishihara} {et~al.}(2010){Ishihara}, {Onaka}, {Kataza}, {Salama},
  {Alfageme}, {Cassatella}, {Cox}, {Garc{\'{\i}}a-Lario}, {Stephenson},
  {Cohen}, {Fujishiro}, {Fujiwara}, {Hasegawa}, {Ita}, {Kim}, {Matsuhara},
  {Murakami}, {M{\"u}ller}, {Nakagawa}, {Ohyama}, {Oyabu}, {Pyo}, {Sakon},
  {Shibai}, {Takita}, {Tanab{\'e}}, {Uemizu}, {Ueno}, {Usui}, {Wada},
  {Watarai}, {Yamamura}, \& {Yamauchi}}]{ishihara10}
{Ishihara}, D., {Onaka}, T., {Kataza}, H., {et~al.} 2010, \aap, 514, A1

\bibitem[{{Jaschek} {et~al.}(1964){Jaschek}, {Conde}, \& {de
  Sierra}}]{jaschek64}
{Jaschek}, C., {Conde}, H., \& {de Sierra}, A.~C. 1964, Observatory
  Astronomical La Plata Series Astronomies, 28, 1

\bibitem[{{Josselin} {et~al.}(1998){Josselin}, {Loup}, {Omont}, {Barnbaum},
  {Nyman}, \& {Sevre}}]{josselin98}
{Josselin}, E., {Loup}, C., {Omont}, A., {et~al.} 1998, \aaps, 129, 45

\bibitem[{{Kataza} {et~al.}(2000){Kataza}, {Okamoto}, {Takubo}, {Onaka},
  {Sako}, {Nakamura}, {Miyata}, \& {Yamashita}}]{kataza00}
{Kataza}, H., {Okamoto}, Y., {Takubo}, S., {et~al.} 2000, in Presented at the
  Society of Photo-Optical Instrumentation Engineers (SPIE) Conference, Vol.
  4008, Society of Photo-Optical Instrumentation Engineers (SPIE) Conference
  Series, ed. {M.~Iye \& A.~F.~Moorwood}, 1144--1152

\bibitem[{{Kennedy}(1996)}]{kennedy96}
{Kennedy}, P.~M. 1996, VizieR Online Data Catalog, 3078, 0

\bibitem[{{Kurucz}(1992)}]{kurucz92}
{Kurucz}, R.~L. 1992, in IAU Symposium, Vol. 149, The Stellar Populations of
  Galaxies, ed. {B.~Barbuy \& A.~Renzini}, 225--+

\bibitem[{{Lecavelier Des Etangs} {et~al.}(1996){Lecavelier Des Etangs},
  {Vidal-Madjar}, \& {Ferlet}}]{lecavelier96}
{Lecavelier Des Etangs}, A., {Vidal-Madjar}, A., \& {Ferlet}, R. 1996, \aap,
  307, 542

\bibitem[{{Liseau} {et~al.}(2010){Liseau}, {Eiroa}, {Fedele}, {Augereau},
  {Olofsson}, {Gonz{\'a}lez}, {Maldonado}, {Montesinos}, {Mora}, {Absil},
  {Ardila}, {Barrado}, {Bayo}, {Beichman}, {Bryden}, {Danchi}, {Del Burgo},
  {Ertel}, {Fridlund}, {Heras}, {Krivov}, {Launhardt}, {Lebreton}, {L{\"o}hne},
  {Marshall}, {Meeus}, {M{\"u}ller}, {Pilbratt}, {Roberge}, {Rodmann},
  {Solano}, {Stapelfeldt}, {Th{\'e}bault}, {White}, \& {Wolf}}]{liseau10}
{Liseau}, R., {Eiroa}, C., {Fedele}, D., {et~al.} 2010, \aap, 518, L132

\bibitem[{{L{\"o}hne} {et~al.}(2012){L{\"o}hne}, {Augereau}, {Ertel},
  {Marshall}, {Eiroa}, {Mora}, {Absil}, {Stapelfeldt}, {Th{\'e}bault}, {Bayo},
  {Del Burgo}, {Danchi}, {Krivov}, {Lebreton}, {Letawe}, {Magain}, {Maldonado},
  {Montesinos}, {Pilbratt}, {White}, \& {Wolf}}]{loehne12}
{L{\"o}hne}, T., {Augereau}, J.-C., {Ertel}, S., {et~al.} 2012, \aap, 537, A110

\bibitem[{{Lord}(1992)}]{lord92}
{Lord}, S.~D. 1992, {A new software tool for computing Earth's atmospheric
  transmission of near- and far-infrared radiation}, Tech. rep.

\bibitem[{{Manoj} {et~al.}(2006){Manoj}, {Bhatt}, {Maheswar}, \&
  {Muneer}}]{manoj06}
{Manoj}, P., {Bhatt}, H.~C., {Maheswar}, G., \& {Muneer}, S. 2006, \apj, 653,
  657

\bibitem[{{Melis} {et~al.}(2010){Melis}, {Zuckerman}, {Rhee}, \&
  {Song}}]{melis10}
{Melis}, C., {Zuckerman}, B., {Rhee}, J.~H., \& {Song}, I. 2010, \apjl, 717,
  L57

\bibitem[{{Melis} {et~al.}(2012){Melis}, {Zuckerman}, {Rhee}, {Song}, {Murphy},
  \& {Bessell}}]{melis12}
{Melis}, C., {Zuckerman}, B., {Rhee}, J.~H., {et~al.} 2012, \nat, 487, 74

\bibitem[{{Meyer} {et~al.}(2008){Meyer}, {Carpenter}, {Mamajek}, {Hillenbrand},
  {Hollenbach}, {Moro-Martin}, {Kim}, {Silverstone}, {Najita}, {Hines},
  {Pascucci}, {Stauffer}, {Bouwman}, \& {Backman}}]{meyer08}
{Meyer}, M.~R., {Carpenter}, J.~M., {Mamajek}, E.~E., {et~al.} 2008, \apjl,
  673, L181

\bibitem[{{Montesinos} {et~al.}(2009){Montesinos}, {Eiroa}, {Mora}, \&
  {Mer{\'{\i}}n}}]{montesinos09}
{Montesinos}, B., {Eiroa}, C., {Mora}, A., \& {Mer{\'{\i}}n}, B. 2009, \aap,
  495, 901

\bibitem[{{Morales} {et~al.}(2009){Morales}, {Werner}, {Bryden}, {Plavchan},
  {Stapelfeldt}, {Rieke}, {Su}, {Beichman}, {Chen}, {Grogan}, {Kenyon},
  {Moro-Martin}, \& {Wolf}}]{morales09}
{Morales}, F.~Y., {Werner}, M.~W., {Bryden}, G., {et~al.} 2009, \apj, 699, 1067

\bibitem[{{Mosser} {et~al.}(2009){Mosser}, {Michel}, {Appourchaux}, {Barban},
  {Baudin}, {Boumier}, {Bruntt}, {Catala}, {Deheuvels}, {Garc{\'{\i}}a},
  {Gaulme}, {Regulo}, {Roxburgh}, {Samadi}, {Verner}, {Auvergne}, {Baglin},
  {Ballot}, {Benomar}, \& {Mathur}}]{mosser09}
{Mosser}, B., {Michel}, E., {Appourchaux}, T., {et~al.} 2009, \aap, 506, 33

\bibitem[{{Mouri} {et~al.}(2011){Mouri}, {Kaneda}, {Ishihara}, {Oyabu},
  {Yamagishi}, {Mori}, {Onaka}, {Wada}, \& {Kataza}}]{mouri11}
{Mouri}, A., {Kaneda}, H., {Ishihara}, D., {et~al.} 2011, \pasp, 123, 561

\bibitem[{{Murakami} {et~al.}(2007){Murakami}, {Baba}, {Barthel}, {Clements},
  {Cohen}, {Doi}, {Enya}, {Figueredo}, {Fujishiro}, {Fujiwara}, {Fujiwara},
  {Garcia-Lario}, {Goto}, {Hasegawa}, {Hibi}, {Hirao}, {Hiromoto}, {Hong},
  {Imai}, {Ishigaki}, {Ishiguro}, {Ishihara}, {Ita}, {Jeong}, {Jeong},
  {Kaneda}, {Kataza}, {Kawada}, {Kawai}, {Kawamura}, {Kessler}, {Kester},
  {Kii}, {Kim}, {Kim}, {Kobayashi}, {Koo}, {Kwon}, {Lee}, {Lorente}, {Makiuti},
  {Matsuhara}, {Matsumoto}, {Matsuo}, {Matsuura}, {M{\"u}ller}, {Murakami},
  {Nagata}, {Nakagawa}, {Naoi}, {Narita}, {Noda}, {Oh}, {Ohnishi}, {Ohyama},
  {Okada}, {Okuda}, {Oliver}, {Onaka}, {Ootsubo}, {Oyabu}, {Pak}, {Park},
  {Pearson}, {Rowan-Robinson}, {Saito}, {Sakon}, {Salama}, {Sato}, {Savage},
  {Serjeant}, {Shibai}, {Shirahata}, {Sohn}, {Suzuki}, {Takagi}, {Takahashi},
  {Tanab{\'e}}, {Takeuchi}, {Takita}, {Thomson}, {Uemizu}, {Ueno}, {Usui},
  {Verdugo}, {Wada}, {Wang}, {Watabe}, {Watarai}, {White}, {Yamamura},
  {Yamauchi}, \& {Yasuda}}]{murakami07}
{Murakami}, H., {Baba}, H., {Barthel}, P., {et~al.} 2007, \pasj, 59, 369

\bibitem[{{Okamoto} {et~al.}(2003){Okamoto}, {Kataza}, {Yamashita}, {Miyata},
  {Sako}, {Takubo}, {Honda}, \& {Onaka}}]{okamoto02}
{Okamoto}, Y.~K., {Kataza}, H., {Yamashita}, T., {et~al.} 2003, in Presented at
  the Society of Photo-Optical Instrumentation Engineers (SPIE) Conference,
  Vol. 4841, Society of Photo-Optical Instrumentation Engineers (SPIE)
  Conference Series, ed. {M.~Iye \& A.~F.~M.~Moorwood}, 169--180

\bibitem[{{Olofsson} {et~al.}(2012){Olofsson}, {Juh{\'a}sz}, {Henning},
  {Mutschke}, {Tamanai}, {Mo{\'o}r}, \& {{\'A}brah{\'a}m}}]{olofsson12}
{Olofsson}, J., {Juh{\'a}sz}, A., {Henning}, T., {et~al.} 2012, \aap, 542, A90

\bibitem[{{Onaka} {et~al.}(2007){Onaka}, {Matsuhara}, {Wada}, {Fujishiro},
  {Fujiwara}, {Ishigaki}, {Ishihara}, {Ita}, {Kataza}, {Kim}, {Matsumoto},
  {Murakami}, {Ohyama}, {Oyabu}, {Sakon}, {Tanab{\'e}}, {Takagi}, {Uemizu},
  {Ueno}, {Usui}, {Watarai}, {Cohen}, {Enya}, {Ootsubo}, {Pearson}, {Takeyama},
  {Yamamuro}, \& {Ikeda}}]{onaka07}
{Onaka}, T., {Matsuhara}, H., {Wada}, T., {et~al.} 2007, \pasj, 59, 401

\bibitem[{{Oudmaijer} {et~al.}(1992){Oudmaijer}, {van der Veen}, {Waters},
  {Trams}, {Waelkens}, \& {Engelsman}}]{oudmaijer92}
{Oudmaijer}, R.~D., {van der Veen}, W.~E.~C.~J., {Waters}, L.~B.~F.~M.,
  {et~al.} 1992, \aaps, 96, 625

\bibitem[{{Pei}(1992)}]{pei92}
{Pei}, Y.~C. 1992, \apj, 395, 130

\bibitem[{{Pilbratt} {et~al.}(2010){Pilbratt}, {Riedinger}, {Passvogel},
  {Crone}, {Doyle}, {Gageur}, {Heras}, {Jewell}, {Metcalfe}, {Ott}, \&
  {Schmidt}}]{pilbratt10}
{Pilbratt}, G.~L., {Riedinger}, J.~R., {Passvogel}, T., {et~al.} 2010, \aap,
  518, L1

\bibitem[{{Price} {et~al.}(2001){Price}, {Egan}, {Carey}, {Mizuno}, \&
  {Kuchar}}]{price01}
{Price}, S.~D., {Egan}, M.~P., {Carey}, S.~J., {Mizuno}, D.~R., \& {Kuchar},
  T.~A. 2001, \aj, 121, 2819

\bibitem[{{Rebull} {et~al.}(2008){Rebull}, {Stapelfeldt}, {Werner}, {Mannings},
  {Chen}, {Stauffer}, {Smith}, {Song}, {Hines}, \& {Low}}]{rebull08}
{Rebull}, L.~M., {Stapelfeldt}, K.~R., {Werner}, M.~W., {et~al.} 2008, \apj,
  681, 1484

\bibitem[{{Rhee} {et~al.}(2007){Rhee}, {Song}, {Zuckerman}, \&
  {McElwain}}]{rhee07}
{Rhee}, J.~H., {Song}, I., {Zuckerman}, B., \& {McElwain}, M. 2007, \apj, 660,
  1556

\bibitem[{{Roeser} \& {Bastian}(1988)}]{roeser88}
{Roeser}, S. \& {Bastian}, U. 1988, \aaps, 74, 449

\bibitem[{{Sako} {et~al.}(2003){Sako}, {Okamoto}, {Kataza}, {Miyata}, {Takubo},
  {Honda}, {Fujiyoshi}, {Onaka}, \& {Yamashita}}]{sako03}
{Sako}, S., {Okamoto}, Y.~K., {Kataza}, H., {et~al.} 2003, \pasp, 115, 1407

\bibitem[{{Sch{\"u}tz} {et~al.}(2009){Sch{\"u}tz}, {Meeus}, {Sterzik}, \&
  {Peeters}}]{schutz09}
{Sch{\"u}tz}, O., {Meeus}, G., {Sterzik}, M.~F., \& {Peeters}, E. 2009, \aap,
  507, 261

\bibitem[{{Sloan} \& {Price}(1998)}]{sloan98}
{Sloan}, G.~C. \& {Price}, S.~D. 1998, \apjs, 119, 141

\bibitem[{{Smith} {et~al.}(2006){Smith}, {Hines}, {Low}, {Gehrz}, {Polomski},
  \& {Woodward}}]{smith06}
{Smith}, P.~S., {Hines}, D.~C., {Low}, F.~J., {et~al.} 2006, \apjl, 644, L125

\bibitem[{{Su} {et~al.}(2006){Su}, {Rieke}, {Stansberry}, {Bryden},
  {Stapelfeldt}, {Trilling}, {Muzerolle}, {Beichman}, {Moro-Martin}, {Hines},
  \& {Werner}}]{su06}
{Su}, K.~Y.~L., {Rieke}, G.~H., {Stansberry}, J.~A., {et~al.} 2006, \apj, 653,
  675

\bibitem[{{Takeda}(2007)}]{takeda07}
{Takeda}, Y. 2007, \pasj, 59, 335

\bibitem[{{Telesco} {et~al.}(1998){Telesco}, {Pina}, {Hanna}, {Julian}, {Hon},
  \& {Kisko}}]{telesco98}
{Telesco}, C.~M., {Pina}, R.~K., {Hanna}, K.~T., {et~al.} 1998, in Presented at
  the Society of Photo-Optical Instrumentation Engineers (SPIE) Conference,
  Vol. 3354, Society of Photo-Optical Instrumentation Engineers (SPIE)
  Conference Series, ed. {A.~M.~Fowler}, 534--544

\bibitem[{{Thompson} {et~al.}(2010){Thompson}, {Smith}, {Stevens}, {Jarvis},
  {Vidal Perez}, {Marshall}, {Dunne}, {Eales}, {White}, {Leeuw}, {Sibthorpe},
  {Baes}, {Gonz{\'a}lez-Solares}, {Scott}, {Vieiria}, {Amblard}, {Auld},
  {Bonfield}, {Burgarella}, {Buttiglione}, {Cava}, {Clements}, {Cooray},
  {Dariush}, {de Zotti}, {Dye}, {Eales}, {Frayer}, {Fritz}, {Gonzalez-Nuevo},
  {Herranz}, {Ibar}, {Ivison}, {Lagache}, {Lopez-Caniego}, {Maddox},
  {Negrello}, {Pascale}, {Pohlen}, {Rigby}, {Rodighiero}, {Samui}, {Serjeant},
  {Temi}, {Valtchanov}, \& {Verma}}]{thompson10}
{Thompson}, M.~A., {Smith}, D.~J.~B., {Stevens}, J.~A., {et~al.} 2010, \aap,
  518, L134

\bibitem[{{Uzpen} {et~al.}(2007){Uzpen}, {Kobulnicky}, {Monson}, {Pierce},
  {Clemens}, {Backman}, {Meade}, {Babler}, {Indebetouw}, {Whitney}, {Watson},
  {Wolfire}, {Benjamin}, {Bracker}, {Bania}, {Cohen}, {Cyganowski}, {Devine},
  {Heitsch}, {Jackson}, {Mathis}, {Mercer}, {Povich}, {Rho}, {Robitaille},
  {Sewilo}, {Stolovy}, {Watson}, {Wolff}, \& {Churchwell}}]{uzpen07}
{Uzpen}, B., {Kobulnicky}, H.~A., {Monson}, A.~J., {et~al.} 2007, \apj, 658,
  1264

\bibitem[{{Winters} {et~al.}(2003){Winters}, {Le Bertre}, {Jeong}, {Nyman}, \&
  {Epchtein}}]{winters03}
{Winters}, J.~M., {Le Bertre}, T., {Jeong}, K.~S., {Nyman}, L., \& {Epchtein},
  N. 2003, \aap, 409, 715

\bibitem[{{Wright} {et~al.}(2003){Wright}, {Egan}, {Kraemer}, \&
  {Price}}]{wright03}
{Wright}, C.~O., {Egan}, M.~P., {Kraemer}, K.~E., \& {Price}, S.~D. 2003, \aj,
  125, 359

\bibitem[{{Wyatt}(2008)}]{wyatt08}
{Wyatt}, M.~C. 2008, \araa, 46, 339

\bibitem[{{Wyatt} {et~al.}(2007{\natexlab{a}}){Wyatt}, {Smith}, {Greaves},
  {Beichman}, {Bryden}, \& {Lisse}}]{wyatt07}
{Wyatt}, M.~C., {Smith}, R., {Greaves}, J.~S., {et~al.} 2007{\natexlab{a}},
  \apj, 658, 569

\bibitem[{{Wyatt} {et~al.}(2007{\natexlab{b}}){Wyatt}, {Smith}, {Su}, {Rieke},
  {Greaves}, {Beichman}, \& {Bryden}}]{wyatt07b}
{Wyatt}, M.~C., {Smith}, R., {Su}, K.~Y.~L., {et~al.} 2007{\natexlab{b}}, \apj,
  663, 365

\bibitem[{{Yamamura} {et~al.}(2010){Yamamura}, {Makiuti}, {Ikeda}, {Fukuda},
  {Oyabu}, {Koga}, \& {White}}]{yamamura10}
{Yamamura}, I., {Makiuti}, S., {Ikeda}, N., {et~al.} 2010, VizieR Online Data
  Catalog, 2298, 0

\bibitem[{{Zuckerman} \& {Webb}(2000)}]{zuckerman00}
{Zuckerman}, B. \& {Webb}, R.~A. 2000, \apj, 535, 959

\end{thebibliography}


\appendix
\section{Logs of Subaru/COMICS and Gemini/T-ReCS Follow-up Observations}
Here we show the logs of follow-up MIR observations of some debris disk candidates
by Subaru/COMICS (Table~\ref{comicsobs}) and Gemini/T-ReCS (Table~\ref{trecsobs}).


\begin{table}
\caption{Log of Subaru/COMICS observations. \label{comicsobs}}
\centering                          
\begin{tabular}{ccccc}        
\hline\hline                 
Object & Filter      & Date & Integ. & Comment \\
       & ($\micron$) & (UT) & (s)    &         \\
\hline                        
HD~15407  & 18.8     & 16 July 2008 &2400 &   \\
5~Lac     & 18.8     & 16 July 2008 & 100 & Standard \\
HD~15407  & 11.7     & 17 July 2008 & 300 &   \\
HD~15407  & 8.8      & 17 July 2008 & 100 &   \\
$\gamma$~Aql & 8.8      & 17 July 2008 &  20 & Standard \\
$\gamma$~Aql & 11.7     & 17 July 2008 &  20 & Standard \\
\hline 
HD~165014 & 18.8     & 16 July 2008 & 400 &   \\
$\gamma$~Aql & 18.8     & 16 July 2008 &  20 & Standard \\
HD~165014 & 11.7     & 17 July 2008 & 200 &   \\
HD~165014 & 8.8      & 17 July 2008 & 100 &   \\
$\epsilon$~Sco & 11.7     & 17 July 2008 &  10 & Standard \\
$\epsilon$~Sco & 8.8      & 17 July 2008 &  50 & Standard \\
\hline 
HD~175726 & 11.7     & 17 July 2008 & 300 &   \\
$\gamma$~Aql & 11.7     & 17 July 2008 &  20 & Standard \\
\hline                                   
\end{tabular}
\end{table}


\begin{table}
\caption{Log of Gemini/T-ReCS observations. \label{trecsobs}}
\centering                          
\begin{tabular}{ccccc}        
\hline\hline                 
Object & Filter      & Date & Integ. & Comment \\
       & ($\micron$) & (UT) & (s)    &         \\
\hline                        
HD~105209 & 18.3     & 10 June 2008 &2172 &   \\
HD~110458 & 18.3     & 10 June 2008 & 116 & Standard \\
HD~105209 & 12.3     & 11 June 2008 & 116 &   \\
HD~105209 & 11.7     & 11 June 2008 & 116 &   \\
HD~105209 & 10.4     & 11 June 2008 & 116 &   \\
HD~105209 & 9.7      & 11 June 2008 & 116 &   \\
HD~105209 & 8.8      & 11 June 2008 & 116 &   \\
HD~110458 & 8.8      & 12 June 2008 &  58 & Standard \\
HD~110458 & 9.7      & 12 June 2008 &  58 & Standard \\
HD~110458 & 10.4     & 12 June 2008 &  58 & Standard \\
HD~110458 & 11.7     & 12 June 2008 &  58 & Standard \\
HD~110458 & 12.3     & 12 June 2008 &  58 & Standard \\
\hline
HD~106797 & 18.3     & 10 June 2008 & 811 &   \\
HD~110458 & 18.3     & 10 June 2008 & 116 & Standard \\
HD~106797 & 12.3     & 12 June 2008 & 174 &   \\
HD~106797 & 11.7     & 12 June 2008 & 116 &   \\
HD~106797 & 10.4     & 12 June 2008 &  58 &   \\
HD~106797 & 9.7      & 12 June 2008 & 174 &   \\
HD~106797 & 8.8      & 12 June 2008 &  58 &   \\
HD~110458 & 8.8      & 12 June 2008 &  58 & Standard \\
HD~110458 & 9.7      & 12 June 2008 &  58 & Standard \\
HD~110458 & 10.4     & 12 June 2008 &  58 & Standard \\
HD~110458 & 11.7     & 12 June 2008 &  58 & Standard \\
HD~110458 & 12.3     & 12 June 2008 &  58 & Standard \\
\hline
HD~110058 & 11.7     & 12 June 2008 & 869 &   \\
HD~110458 & 11.7     & 12 June 2008 &  58 & Standard \\
HD~110058 & 18.3     & 26 June 2008 & 260 &   \\
HD~110458 & 18.3     & 26 June 2008 & 116 & Standard \\
\hline                                   
\end{tabular}
\end{table}

\end{document}